\documentclass{article}
\usepackage{emulateapj,epsfig}

\def\ltsima{$\; \buildrel < \over \sim \;$}
\def\simlt{\lower.5ex\hbox{\ltsima}}
\def\gtsima{$\; \buildrel > \over \sim \;$}
\def\simgt{\lower.5ex\hbox{\gtsima}}

\slugcomment{To appear in ApJ}

\makeatletter

\newenvironment{inlinefigure}{%
\def\@captype{figure}%
\noindent\begin{minipage}{0.999\linewidth}\begin{center}}
{\end{center}\end{minipage}\smallskip}
\makeatother

\begin{document}

\title{Early Structure Formation and Reionization in a Cosmological Model 
with a Running Primordial Power Spectrum}

\author{Naoki Yoshida$^{1,2}$, Aaron Sokasian$^{1}$, Lars Hernquist$^{1}$}
\affil{$^1$Harvard-Smithsonian Center for Astrophysics, 60 Garden Street,
Cambridge MA02138}
\affil{$^2$National Astronomical Observatory Japan, Mitaka, Tokyo 181-8588, Japan}
\and
\author{Volker Springel$^{3}$}
\affil{$^3$Max-Planck-Institut f\"ur Astrophysik, Karl-Schwarzschild-str. 1, D-85740
Garching bei M\"unchen, Germany}

\begin{abstract}

We study high redshift structure formation and reionization in a
$\Lambda$CDM universe under the assumption that the spectral power
index of primordial density fluctuations is a function of length
scale. We adopt a particular formulation of the ``running'' spectral
index (RSI) model as suggested by the combined analysis of the recent
WMAP data and two other large-scale structure observations. We
carry out high resolution cosmological simulations and use them to
study the formation of primordial gas clouds where the first stars
are likely to form.

While early structure forms hierarchically in the RSI model, quite 
similarly to the standard power-law $\Lambda$CDM model, the reduced
power on small scales causes a considerable delay in the formation
epoch of low mass ($\sim 10^6 M_{\odot}$) ``mini-halos'' compared to
the $\Lambda$CDM model.  The abundance of primordial star-forming gas
clouds in such halos also differs by more than an order of magnitude
at $z>15$ between the two models.  The extremely small number of gas
clouds in the RSI model indicates that reionization is initiated later
than $z<15$, generally resulting in a smaller total Thomson optical
depth than in the $\Lambda$CDM model.

By carrying out radiative transfer calculations, we also study
reionization by stellar populations formed in galaxies.  We show that,
in order to reionize the Universe by $z \sim 7$, the escape fraction
of ultraviolet photons from galaxies in the RSI model must be as high
as $~0.6$ throughout the redshift range $5<z<18$ for a stellar
population similar to that of the local Universe.  Even with a
top-heavy intial mass function representing an early population of
massive stars and/or an extraordinarily high photon emission rate from
galaxies, the total optical depth can only be as large as $\tau_{e}
\sim 0.1$ for reasonable models of early star-formation. The RSI model
is thus in conflict with the large Thomson optical depth inferred by
the WMAP satellite.

\end{abstract}


\section{Introduction}

The origin of matter density fluctuations in the Universe is one of
the fundamental problems in cosmology.  The so-called standard theory
of structure formation posits that the present-day clumpy appearance
of the Universe developed through gravitational amplification of an
initially very smooth, but perturbed matter distribution.  Recent
observations provide a consistent picture of the
large-scale matter distribution in the early Universe; fluctuations
arise from adiabatic perturbations whose statistics are described by
a Gaussian field, as predicted by
popular inflationary theories.  The
inflationary models also predict a scale-invariant power
spectrum of primordial density fluctuations. 
In fact, the generic inflationary prediction is {\it nearly but
not precisely} scale-invariant density fluctuations; i.e.  the
fluctuation power spectrum scales as $P(k)\propto k^n$ with $n\approx
1$ (see Lyth \& Riotto 1999 for a recent review).  The tilt, or the
exact value for $n$, has attracted considerable attention in theories
of the physics of the early Universe (Covi \& Lyth 1999), in astronomical
observations (Croft et al. 2002; Hannestad et al. 2002; Seljak,
McDonald \& Makarov 2003) and in the context of structure formation
(Zentner \& Bullock 2002). 

The spectral index of the primordial power, $n$, may even be 
a function of length scale.
This ``running'' of the spectral index modulates
the relative amplitudes of density fluctuations on large to small
scales. Interestingly, the combined
analysis of the first-year Wilkinson Microwave Anisotropy Probe
(WMAP) data, the 2dF galaxy redshift survey, and Lyman-$\alpha$
forest observations favors a cosmological model with a ``running'' of
the primordial power spectrum (Spergel et al. 2003; Peiris et
al. 2003).  Although the WMAP data alone gives a best-fit power law
$\Lambda$CDM model with the spectral index $n\approx 1$ and the
fluctuation amplitude $\sigma_8 = 0.9$, the Lyman-$\alpha$ forests
observations (Croft et al. 1998, 2002; McDonald et al. 2000; 
Gnedin \& Hamilton 2002) consistently
favor a low-amplitude model on scales $1 \lesssim k\;[{\rm Mpc}^{-1}]
\lesssim 10$, which can be reconciled if the primordial power spectrum
has a mild tilt ($n<1$) or a negative running ($dn/d\ln k < 0$).
For a given normalization on a large scale, models with
negative running predict progressively reduced linear power on small
scales (Kosowsky \& Turner 1995; Hannestad et al. 2002).

Suppressing linear power on small, galactic to sub-galactic scales may
offer a resolution to the often-claimed problems of Cold
Dark Matter (CDM) models (e.g. Moore et al. 1999).  
It could alleviate apparent discrepancies
between the observed structure of dark halos and predictions from CDM
models (Zentner \& Bullock 2003). On the other hand,
too much reduction of small-scale power may result in a conflict with
other observations.  Dalal \& Kochanek (2002) and Chiba (2002) argue
that the abundance of dark matter substructure in several galactic
lens systems inferred from observed flux anomalies is consistent
with predictions from the standard $\Lambda$CDM model (see however
Bullock \& Zentner 2002 for an attempt to constrain some variant
models).  In models with reduced small scale power, nonlinear objects
with characteristic mass $10^6-10^9 M_{\odot}$ form late, and hence
reionization due to photons from stellar sources formed in low
mass systems is expected to occur late (Somerville, Bullock \& Livio
2003).  To the contrary, the measurement of TE polarization by WMAP
suggests a large Thomson optical depth $\tau_{e}=0.17\pm 0.04$,
implying that reionization took place as early as $z_{\rm reion}\sim
17$ (Kogut et al. 2003).  

 Somerville et al. (2003) argue that models
like the RSI one we consider here are inconsistent with the inferred
early reionization epoch.  Yoshida et al. (2003b) studied early
structure formation in a warm dark matter model and showed that
suppressing the linear power even on very small scales $(k > 100$
Mpc$^{-1}$) can considerably affect the formation of high redshift
star-forming gas clouds, making early reionization by massive,
metal-free stars unlikely to occur early on.  Chiu, Fan \& Ostriker
(2003) argue that a large optical depth $\tau_{e} > 0.1$ may require
exotic radiation sources other than a normal stellar
population. Avelino \& Liddle (2003) also conclude that running of
the primordial power systematically delays the reionization epoch.
Overall, these theoretical studies suggest that, {\it if} the large
Thomson optical depth observed by the WMAP satellite is confirmed, the
standard power-law ($n\approx 1$) $\Lambda$CDM model is favored over
other models with reduced small scale power.  Thus, the WMAP results
appear to suggest a somewhat contradictory cosmological model. Since
the major reason to favor power-law $\Lambda$CDM in this context
is the inferred early epoch of reionization, it is clearly important
to study and compare quantitatively details of the 
reionization process in
the two cosmologies, the RSI model and a power law $\Lambda$CDM universe.

In the present paper, we study early structure formation in RSI and
$\Lambda$CDM universes using cosmological $N$-body/hydrodynamic
simulations.  By combining these dynamical models with radiative
transfer calculations, we compute the epoch of reionization by various
stellar sources in the two cosmologies and compare the Thomson optical
depth to that measured by the WMAP satellite.

In particular, at high redshifts, $z\simgt 15$, we consider the
implications of an early population of massive, metal-free stars (the
``first stars''), which have been hypothesized to form in
``mini-halos'' ($\sim 10^6 M_{\odot}$).  We refer to the sites where
these stars can form as ``primordial gas clouds;'' i.e. dense regions
of cold gas which accumulate through the action of cooling by
molecular hydrogen.  In order to identify these primordial gas clouds,
our simulations follow the non-equilibrium chemical reactions of a
chemically pristine gas and molecular hydrogen cooling.  In an earlier
study (Yoshida et al. 2003a), we found that non-equilibrium effects
associated with dynamical heating play a critical role in regulating
the supply of molecular gas, contrary to the assumptions commonly
made in semi-analytic studies of this process.  Hence, it is
necessary to follow the dynamics of structure formation 
explicitly to reliably estimate the abundance of early stars.
It is also important to specify the source location accurately 
when carrying out radiative transfer caculations. Unlike semi-analytic
methods, direct hydrodynamic simulations enable us to locate
plausible star-forming sites robustly in a cosmological volume.
As we discuss later in the present paper, the distribution of 
the sources strongly affects the overall topology of the ionized regions. 

We show that the reduced small scale power in the RSI model causes a
considerable delay in the formation epoch of low mass halos and the
primordial gas clouds within them compared to a $\Lambda$CDM
counterpart.  In view of this result, it appears unlikely that massive
stars in mini-halos can contribute significantly to reionization in
cosmologies with a running spectral index, at least for those with
parameters similar to what is inferred from the joint 
WMAPext + 2dFGRS + Lyman-$\alpha$ forest analysis (Spergel et al. 2003).
To study the impact of ordinary (``Population II'')
stars formed in galaxies and protogalaxies, we carry out ray tracing
calculations of reionization by stellar sources.  For an initial mass
function (IMF) similar to that in local galaxies, reionization can be
completed at sufficiently early times to account for the
Lyman-$\alpha$ optical depths measured in quasars at $z>6$, but in the
case of the RSI model only if a large fraction of ionizing photons are
able to escape from galaxies.  Furthermore, because reionization
occurs late in this context, the Thomson optical depth is too small
to be compatible with that suggested by the WMAP measurements.

Finally, we consider the possibility that the production rate of
ionizing photons could be boosted in galaxies and protogalaxies with a
top-heavy IMF.  In this case, Thomson optical depths approaching that
favored by the WMAP analysis are possible, but only if the parameters
describing these stars are pushed to extreme values.

The paper is organized as follows. We first describe the cosmological
simulations we use to study the abundance of high-redshift
star-forming systems in \S 2. We present the simulation results
in \S 3.  In \S 4, we discuss the possible effect of the finite
size of our simulation volume.  In \S 5, the abundance of
primordial gas clouds and implications are described. There, we
also give a
brief description of the basic relation between reionization epoch and
the total Thomson optical depth.  In \S 6 we show
the results of radiative transfer calculations in which we consider
ordinary stellar populations in galaxies and protogalaxies as
radiation sources.  A summary and discussion are given in \S 7.

\begin{inlinefigure}
\resizebox{8cm}{!}{\includegraphics{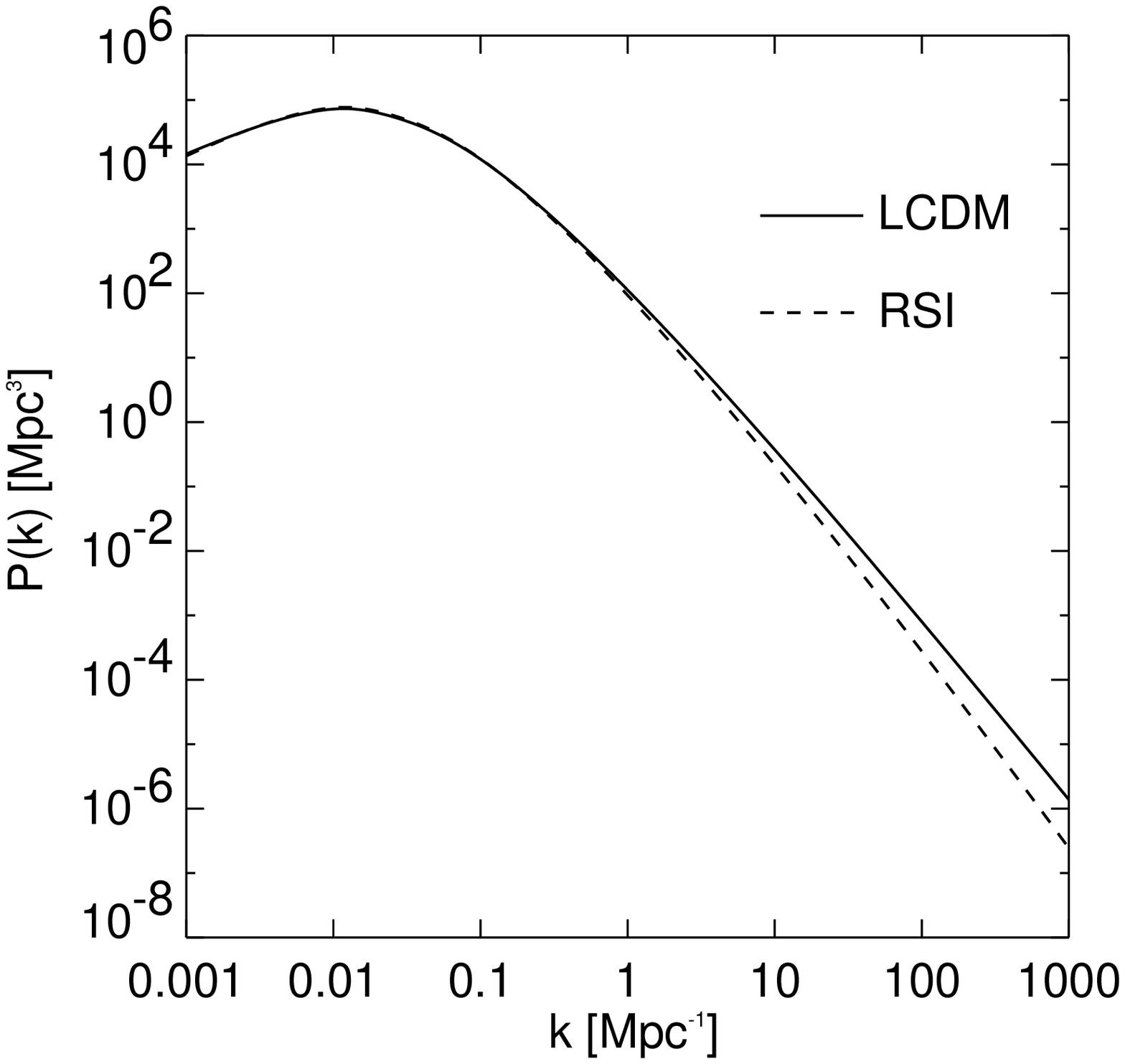}}
\resizebox{8cm}{!}{\includegraphics{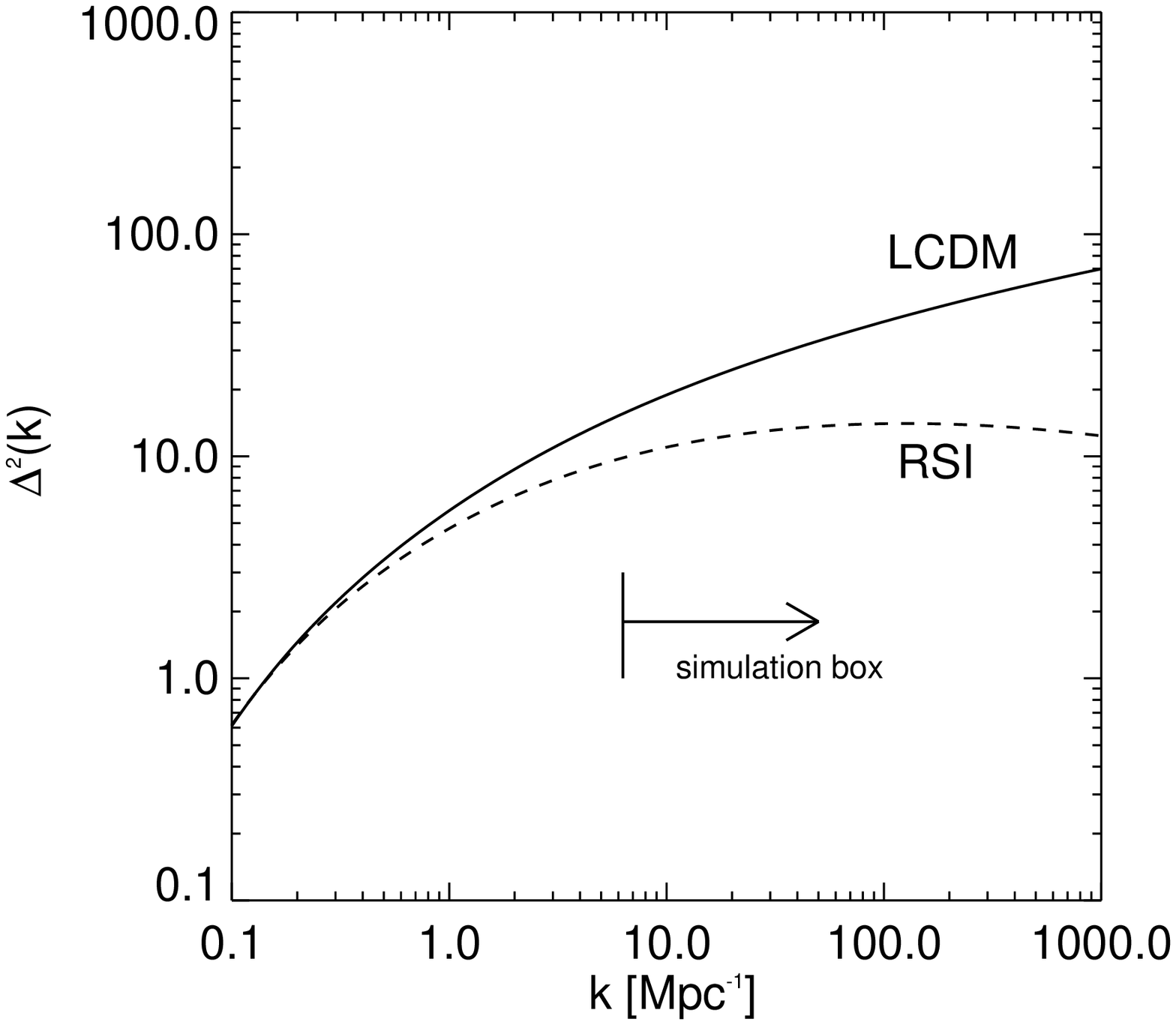}}
\caption{The linear power spectra for the power law $\Lambda$CDM model (solid line)
and for the RSI model (dashed line). In the bottom panel we plot the variance
$\Delta^2 (k)$ linearly extrapolated to the present epoch for the two
models. 
Note the difference in the plotted range in $k$ between the top and the bottom
panels.
The arrow in the bottom panel indicates the
length scales our simulations probe. \label{powspec}}
\end{inlinefigure}

\section{The $N$-body/hydrodynamic simulations}

We use the parallel Tree-PM solver GADGET2 combined with smoothed
particle hydrodynamics (SPH) in our simulations.  Our treatment of SPH
employs the conservative entropy formulation of Springel \& Hernquist
(2002), which offers several distinct advantages over previous
versions of SPH.  In particular, because the energy equation is
written with the entropy as the independent thermodynamic variable, as
opposed to the thermal energy, the '$p{\rm d}V$' term is not evaluated
explicitly, reducing noise from smoothed estimates of e.g. the
density.  By including terms involving derivatives of the density with
respect to the particle smoothing lengths, this approach explicitly
conserves entropy (in regions without shocks), even when smoothing
lengths evolve adaptively, avoiding the problems noted by
e.g. Hernquist (1993).  Furthermore, this formulation moderates the
overcooling problem present in earlier formulations of SPH 
(Yoshida et al. 2002; see also Pearce et al. 1999, Croft et al. 2001).

The simulations follow the non-equilibrium reactions of nine chemical
species (e$^{-}$, H, H$^+$, He, He$^{+}$, He$^{++}$, H$_{2}$,
H$_{2}^{+}$, H$^{-}$) using the reaction coefficients compiled by Abel
et al. (1997). We use the cooling rate of Galli \& Palla (1998) for
molecular hydrogen cooling. Further simulation details are found in
Yoshida et al.  (2003a,b).  The
simulations employ $2\times 324^3$ particles in a comoving volume of 1
Mpc on a side.  We work with $\Lambda$-dominated cosmologies with
matter density $\Omega_{\rm m}=0.3$, cosmological constant
$\Omega_{\Lambda}=0.7$ and the Hubble constant at the present time
$h=0.7$ in units of $100$km s$^{-1}$Mpc$^{-1}$.  We set the baryon
density to $\Omega_{\rm b}=0.04$ and the normalization parameter
to $\sigma_8 =0.9$.  The transfer functions are computed by the Boltzmann
code of Sugiyama (1995) for the adopted cosmology and initial
conditions are realized as described in Yoshida, Sugiyama \&
Hernquist (2003).  The primordial
power spectrum for the $\Lambda$CDM case is assumed to be
$P(k)\propto k$, whereas that of the RSI model is given by
$P(k)\propto k^{n_{\rm s}}$ with
\begin{equation}
n_{\rm s}=n_{\rm s}(k_0)
+\frac{1}{2}\frac{{\rm d}n_{\rm s}}{{\rm d}\ln k}\ln \left(\frac{k}{k_0}\right).
\end{equation}
Specifically, we choose $k_0$=0.05 Mpc$^{-1}$, $n_{\rm s}(k_0)$=0.93,
and $dn_{s}/d\ln k$=-0.03, as indicated by the WMAPext + 2dFGRS +
Lyman-$\alpha$ analysis (Spergel et al. 2003). Note that we adopt the
same normalization, $\sigma_8 =0.9$, for both the power law
$\Lambda$CDM model and the Running Spectral Index model.
Hereafter, we refer to the conventional power law $\Lambda$CDM model
simply as the ``$\Lambda$CDM model,'' and to the other as the ``RSI model''.

Figure \ref{powspec} shows the linear power spectra $P(k)$ for the two
models.  In the bottom panel, we plot the variance per unit
logarithmic interval in $k$, $\Delta^2 (k) = k^3 P(k)/(2\pi^2)$, to
better describe the difference between the two models in the range
$0.1 < k [{\rm Mpc}^{-1}] < 1000$.  Since the input power spectra are
normalized at a length scale larger than the simulation boxsize, the
amplitudes of the density fluctuations realized in the RSI simulation
are smaller than in the $\Lambda$CDM simulation on all length
scales.  Hence, the density fluctuations on large scales ($k \sim
10$) in the RSI model should evolve similarly to those in
$\Lambda$CDM but with a slight delay in time.  We note that, while the
boxsize of our simulations is sufficiently large to ensure that the
fundamental fluctuation mode in the simulations remains in the linear
regime at the early epochs we simulate, it may not sample a {\it fair}
volume of the Universe. We will discuss finite boxsize effects in
\S 4.

Figure \ref{picture} shows the projected gas distribution at $z=22$
(top panels) and at $z=17$ (bottom panels) for the two models.  We
choose these output redshifts such that the density fluctuation on the
largest length scale in the RSI model at $z=17$ resembles that
of the $\Lambda$CDM model at $z=22$.  As expected, the RSI model
appears as a {\it delayed} $\Lambda$CDM model on these scales.  Also,
rich small-scale structure develops early on only in the $\Lambda$CDM
model, as can be inferred from the linear power spectrum (Figure
\ref{powspec}).

\begin{inlinefigure}
\resizebox{8.6cm}{!}{\includegraphics{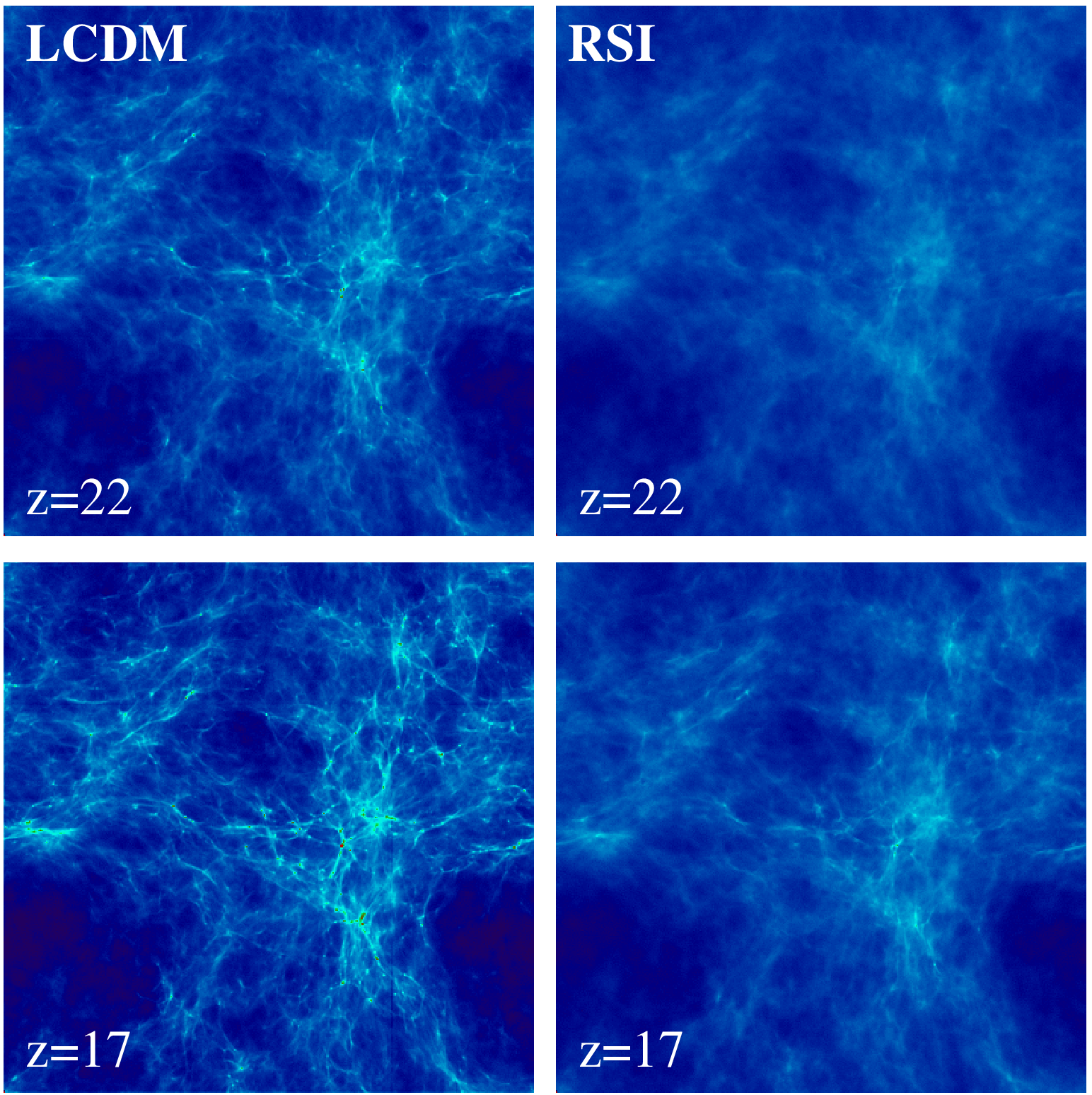}}
\caption{The projected gas density distribution at $z=22$ (top
panels) and $z=17$ (bottom panels) for the $\Lambda$CDM (left)
and RSI (right) models. \label{picture}}
\end{inlinefigure}

\section{Halo abundance}

Since the typical masses of the halos that host early primordial gas
clouds are $5\times 10^5 - 10^7 M_{\odot}$ (Tegmark et al 1997;
Machacek et al. 2001; Yoshida et al. 2003a), the abundance of such
small mass halos is an important quantity to compare. In order to
quantify the difference between the two models, we measure the mass
function of dark matter halos.  Whereas locating dark halos in
pure $N$-body simulations is relatively simple using conventional
techniques such as the friends-of-friends (FOF) algorithm, there are a few
subtleties in identifying and assigning properties to halos in
$N$-body/hydrodynamic simulations. 
Since our main objective is to compare the halo abundance between the
two models, we avoid ambiguities by adopting a simple procedure, as
follows.  

We locate halos by
running a FOF groupfinder with linking parameter $b=0.2$ on the dark
matter particles in our simulations.  We then assign the total mass to
each group as the number of linked member particles multiplied by the
scaled particle mass $m_{\rm part}=m_{\rm dm}+m_{\rm gas}$, where
$m_{\rm dm}$ is the dark matter particle mass and $m_{\rm gas}$ is the
gas particle mass.  We compute the Press-Schechter (PS) mass function
to which we input the linear power spectrum for the dark matter
component. (Note that, for our simulations, separate transfer
functions are used for the dark matter and gas components in
setting-up the initial conditions [for details, see Yoshida, Sugiyama
\& Hernquist 2003].)  We compare the mass function of the halos
identified in this manner at $z=17$ and $z=22$ in Figure
\ref{mfz17}\footnote{Only in Figure \ref{mfz17} and Figure \ref{mfz7},
we use units that include the Hubble constant $h$, to be consistent
with those commonly used; $h^{-1}M_{\odot}$ for mass and $(h^{-1}{\rm
Mpc})^{-3}$ for abundance.}.
For both models, the
measured mass function is overall well fitted by the PS mass function
(the solid line for the $\Lambda$CDM model and the dashed line for the
RSI model). Again, the RSI model at $z=17$ resembles the $\Lambda$CDM
model at $z=22$ with the abundance of low-mass halos being appreciably small.

\begin{inlinefigure}
\resizebox{8.6cm}{!}{\includegraphics{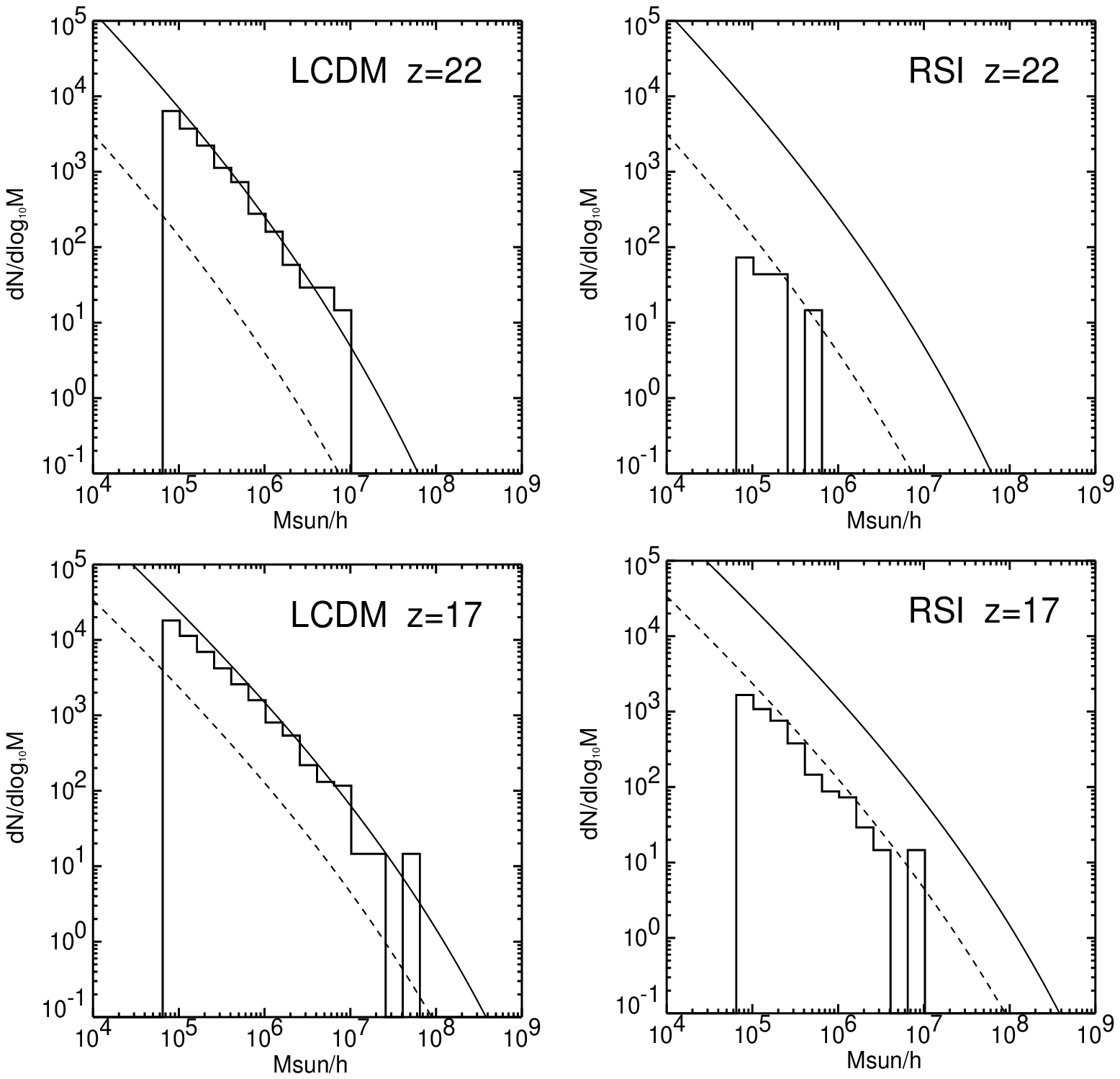}}
\caption{The mass function for the $\Lambda$CDM model (left) 
and for the RSI model (right) at $z=17$ and $z=22$.  
The solid line is the Press-Schechter mass function for 
the $\Lambda$CDM model,
and the dashed line is that for the RSI model. \label{mfz17}}
\end{inlinefigure}

The halo abundance in the RSI model appears to deviate slightly from
the PS mass function. Although a similar feature is also seen in the
$\Lambda$CDM model, the deviation from the PS mass function is more clearly
noticeable for the RSI model.  We discuss this issue together with
finite boxsize effects in the next section.

\section{Mass variance}

It is well known that structure in CDM models grows hierarchically;
i.e., smaller mass objects form earlier and merge to form bigger
objects.  While this can be easily inferred from the variance
$\Delta^2 (k)$ for the $\Lambda$CDM model, which has larger amplitudes
on smaller length scales (see Figure \ref{powspec}), the situation is
less clear for the RSI model.  As Figure \ref{powspec} shows,
the variance in the RSI model appears approximately constant over a
range of scales $10 < k [{\rm Mpc}^{-1}] < 1000$, indicating that nonlinear
objects with widely different masses could form nearly at the same epoch. It is
also worth mentioning that it is not clear whether the Press-Schechter
theory can be used for such models.  For the comparison shown in
Figure \ref{mfz17}, we calculated the PS mass function by formally
inputting the linear power spectrum for the RSI model.  Although the
agreement shown in Figure \ref{mfz17} is reasonably good, it is {\it
not} trivially expected.  The flatness of the variance of the RSI
model deserves particular attention also for another reason; the
hierarchical nature of the formation of dark matter halos has a strong
influence on the formation of primordial gas clouds (Yoshida et
al. 2003a).  Hence it is important to verify whether or not small mass
($\sim 10^6 M_{\odot}$) nonlinear objects in the RSI model grow
hierarchically early on.  For this purpose, it is more appropriate to
work with a properly defined mass variance rather than the power
spectrum.

\begin{inlinefigure}
\resizebox{8cm}{!}{\includegraphics{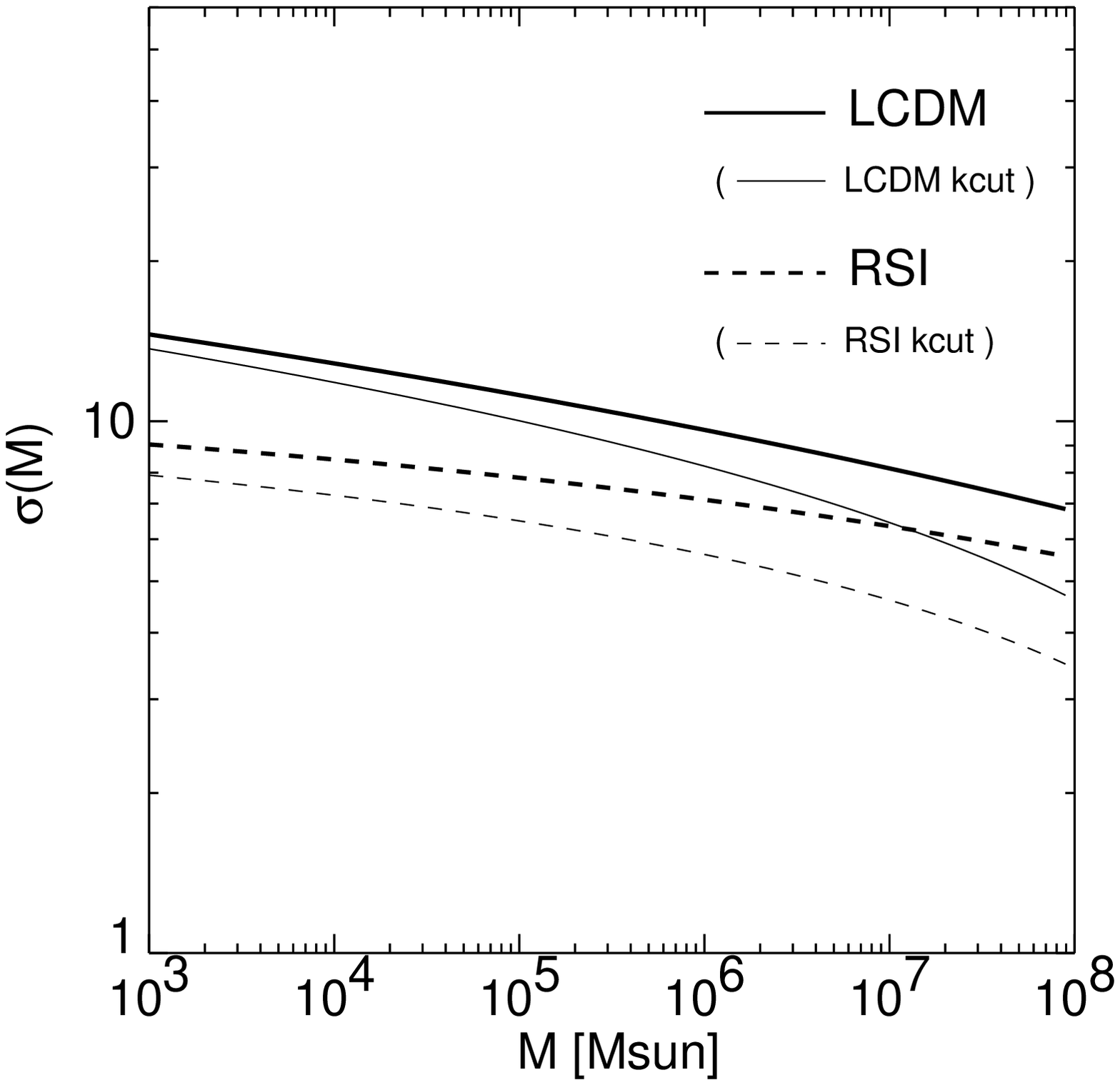}}
\caption{The mass variance computed from
the power spectra linearly extrapolated to the present epoch. 
The thick solid line is for the $\Lambda$CDM model,
and the thick dashed line is for the RSI model.
The thin lines show the
effective mass variance as defined in equation (\ref{sig_eff})
for each model. \label{sigma}}
\end{inlinefigure}
\\
We compute the mass variance as
\begin{equation}
\sigma^2 (M)=\frac{1}{2\pi^2}\int P(k) W^2(kR) k^2 {\rm d}k,
\end{equation}
where the top-hat window function is given by $W(x)=3(\sin (x)/x^3 -
\cos (x)/x^2)$.  Figure \ref{sigma} shows $\sigma (M)$ for the two
models.  In the plotted range, the mass variance in the RSI model
monotonically increases toward smaller mass scales, indicating that
less massive objects form at earlier epochs; i.e., structure formation
is still expected to be ``bottom-up''.  Note that the RSI model
adopted in the present
paper predicts an effective slope of the power spectrum smaller 
than -3 for $k\gg 200$. On very small length
scales ($k > 1000$), or equivalently on low mass scales ($M < 1000
M_{\odot}$), structure formation should then proceed differently from the
``bottom-up'' picture.  
Owing to the flattening of the variance, 
the characteristic mass scale of the first non-linear objects may not 
be well defined.
Halos with masses $10-1000 M_{\odot}$ should collapse first and
larger objects will be formed through mergers of these small halos.  
Simulations with a substantially higher
mass resolution 
are needed to reveal how halos with such very low masses form,
although it will not be important for the formation of halos with masses
$\sim 10^6 M_{\odot}$ that are crucial for primordial gas
cloud formation.

Let us now examine the consequences of the finite box size of our
simulations. We quantify the effect by computing an effective variance
\begin{equation}
\sigma_{\rm eff}^2 (M)=\frac{1}{2\pi^2}\int_{k_{\rm min}} P(k) W^2(kR)
k^2 {\rm d}k,
\label{sig_eff}
\end{equation}
where we set the lower boundary $k_{\rm min}$ to exclude the contribution
from large-scale density fluctuations which are missing in our
simulations.  We simply set $k_{\rm min} =2\pi/L_{\rm box}$ with
$L_{\rm box}$ being the simulation box sidelength.  The thin lines in
Figure \ref{sigma} are the effective mass variance computed in this
manner. For both the models, $\sigma_{\rm eff} (M)$ is 

\begin{inlinefigure}
\resizebox{8cm}{!}{\includegraphics{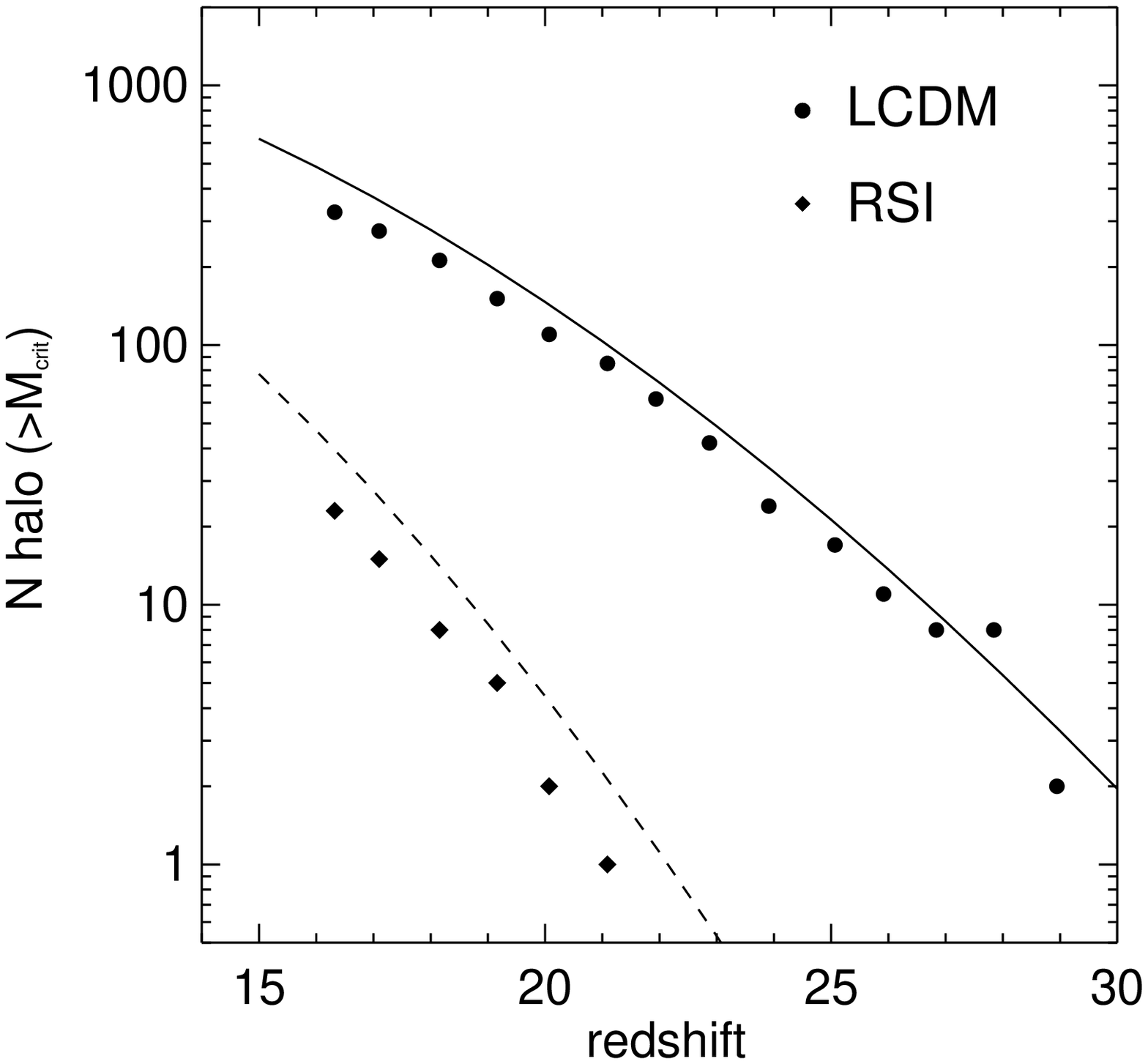}}
\caption{The number of halos with mass greater than $7\times 10^5 M_{\odot}$
per cubic mega-parsec (comoving) volume.
The solid line is the analytic estimate obtained from the Press-Schechter mass function
for the $\Lambda$CDM model,
and the dashed line is for the RSI model. \label{nhalo}}
\end{inlinefigure}
\\
slightly
smaller than $\sigma(M)$ on the relevant mass scales ($\sim 10^5-10^6
M_{\odot}$), and thus a systematic delay in the characteristic
formation epoch (or a slightly lower abundance at a given epoch) is
expected. This may account for the small discrepancy in the mass
function plotted in Figure \ref{mfz17}.  Since we use linearly
extrapolated power spectra to compute $\sigma (M)$ and $\sigma_{\rm
eff} (M)$, the precise effect of a finite boxsize remains still
uncertain.  The apparent difference shown in Figure \ref{sigma} should
be interpreted as {\it qualitatively} representing the finite boxsize
effect.  Note that the RSI model is more affected by the finite
boxsize because the relative fractional power on large scales is
bigger than in the $\Lambda$CDM model (see Figure
\ref{powspec}). Systematic studies using larger box simulations are
clearly needed to address this issue further.

By constructing a halo merger history using 30 simulation outputs
dumped frequently between $17<z<25$, we also verified that almost all
the halos that have masses greater than $7\times 10^5 M_{\odot}$ at
$z=17$ grew monotonically in mass by repeated mergers and
continuous accretion of smaller mass elements, quite similarly to what is
found in the $\Lambda$CDM model at somewhat higher redshifts.  As long
as we are concerned with the abundance of low-mass halos, the
results of our small box simulations appear to be in reasonable
agreement with the Press-Schechter mass function.  In Figure
\ref{nhalo} we show the number of halos with masses greater than $M_{\rm
crit}=7\times 10^5 M_{\odot}$.  We chose the threshold mass $M_{\rm
crit}$ based on the results of Yoshida et al. (2003a) for star-forming
systems. Although the incomplete sampling of the halo mass function
owing to the finite boxsize is appreciable at high redshift, $z>27$ for
the $\Lambda$CDM model and $z>20$ for the RSI model, overall the halo
abundance is in agreement with the PS prediction. Thus, we conclude
that the statistics we discuss in the present paper, such as the
abundance of primordial gas clouds, are not significantly affected by
the finite simulation boxsize. The apparent discrepancy in Figure
\ref{nhalo} may be partly due to some inaccuracies of the analytic
mass function itself (see Jenkins et al. 2001; Reed et al. 2003),
although the simulation results of Jang-Condell \& Hernquist (2001)
indicate that the PS mass function is a good approximation on
these mass scales.  An
important point to note is that both the analytic estimate and the
simulation results suggest that the number of of halos with mass $\sim
10^6 M_{\odot}$, which are crucial for the primordial gas cloud
formation, is an order of magnitude smaller in the RSI model.

Finally, we mention that a promising way of compensating for the missing
large-scale power would be to employ the mode-adding procedure
developed by Tormen \& Bertschinger (1996) and by Cole (1997).  The
method can effectively extend the dynamic range of cosmological
simulations, and thus will be useful for studies of early structure
formation in a cosmological context.

\section{Contribution of the ``first stars'' to reionization}

Numerical simulations (e.g. Abel, Bryan \& Norman 2002; Bromm, Coppi
\& Larson 2002) suggest that the first stars in the Universe were
unusually massive and formed in ``mini-halos'' at redshifts $z>20$, as
a consequence of cooling by molecular hydrogen.  It is customary to
refer to early, massive stars as Population III.  In what follows, we
will distinguish between massive stars which may have formed in
mini-halos by molecular cooling (the ``first stars'') from those which
could have originated in larger halos where gas can cool by atomic
processes.  While the nature of the ``first stars'' has been studied
in some detail, the stellar populations expected in slightly more
massive halos (the ``first galaxies'') is unclear.  With modest
uncertainty, we can quantify the importance of the former to
reionization, but numerous assumptions are required to account for
``non-standard'' IMFs in early protogalaxies, as we discuss later
in the paper.

\subsection{Primordial gas cloud formation}

Owing to the delayed formation of ``mini-halos'' in the RSI model, the
formation of primordial gas clouds is also expected to occur late.
It is in these regions that significant quantities of molecular gas
can accumulate, leading to the formation of massive, metal-free stars
of the type modeled by e.g. Abel, Bryan \& Norman (2002) and Bromm,
Coppi \& Larson (2002).  We emphasize that non-equilibrium processes
not typically accounted for in semi-analytic estimates, like the
dynamical heating effect identified by Yoshida et al. (2003a),
significantly influence the abundance of these star-forming regions.

In Figure \ref{ncloud} we plot the number of gas clouds identified in
the two simulations.  We define groups of cold ($T<500$K), dense
($n_{\rm H} > 500$cm$^{-3}$) gas particles as ``gas clouds''.  We
locate the gas clouds by running a FOF groupfinder to the simulation
gas particles with a small linking parameter $b=0.05$. We then discard
from the groups gas particles that do not satisfy the above
conditions.  Figure \ref{ncloud} clearly shows that the total number
of gas clouds in the simulated volume differs by more than an order of
magnitude between the two models in the redshift range plotted. At
$z=17$, we identified 66 gas clouds in the CDM model, whereas there is
only one gas cloud found in the RSI model.  We continued the RSI run
down to $z=15$ to see if there is a rapid increase in the number of
gas clouds. Although the increase in the number of gas clouds at
$z\sim 14-17$ appears large, the total number is still much smaller
than that of the $\Lambda$CDM model.  
As we argued in section 4, some halos grow very rapidly in the RSI
model, which causes considerable perturbation to the formation
and the growth of the primordial gas clouds within them. 
This may partly explain the extremely small number of
gas clouds at $z\sim 15$ in the RSI model.
It is interesting that the number of
the gas clouds in the RSI model is even {\it smaller} than, or at most
comparable to, the warm dark matter case we studied in Yoshida et al.
(2003b).  For the specific model

\begin{inlinefigure}
\resizebox{8cm}{!}{\includegraphics{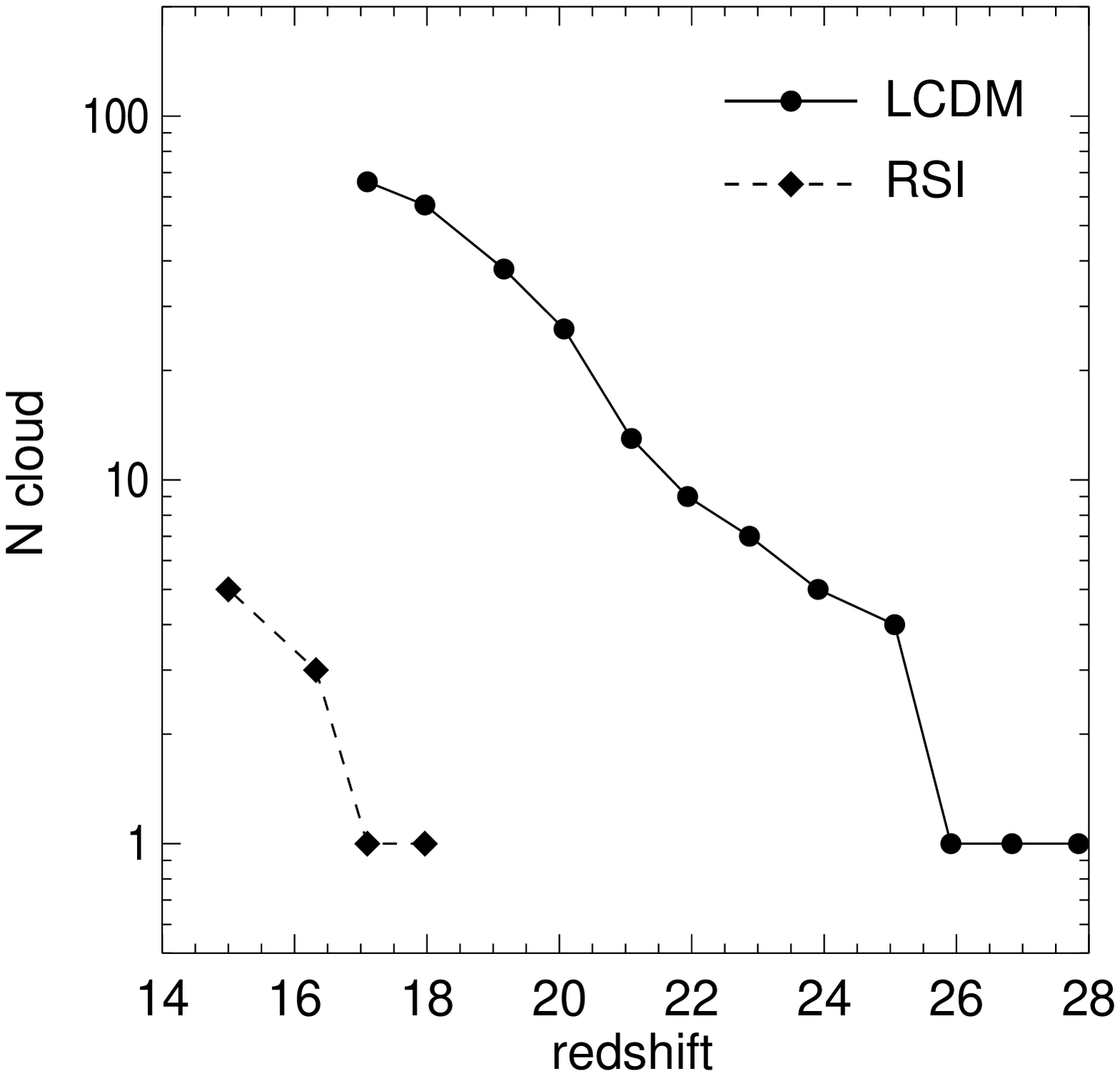}}
\caption{The number of primordial ``gas clouds'' where massive,
metal free 
stars can form by molecular hydrogen cooling. \label{ncloud}}
\end{inlinefigure}
\\
employed in Yoshida et al. (2003b)  that assumes a single
massive ($300 M_{\odot}$), metal-free star is formed per gas cloud,
reionization by the ``first stars'' is not completed at $z >
15$ in the RSI model.  In fact, with the very small number of sources, the
ionized volume fraction rises only up to a few percent by $z=15$.
This is clearly in disagreement with the WMAP result of $z_{\rm
reion}\sim 17$. Since the estimated uncertainty in the WMAP result of
the reionization epoch is rather large (Kogut et al. 2003), it still
remains to be seen whether the particular RSI model we consider is
compatible with the WMAP data.  Nevertheless, it is clear that the
running of the power spectrum is {\it not} favored in terms of early
reionization.

\subsection{Reionization epoch and the Thomson optical depth}

The differential optical depth to Thomson scattering in a
small redshift interval ${\rm d}z$ is
\begin{equation}
{\rm d}\tau_{e}=\sigma_{\rm T}\,n_{e}(z)\,c
\frac{{\rm d}t}{{\rm d}z}\,{\rm d}z,
\end{equation}
where $\sigma_{\rm T}=6.65\times 10^{-25} {\rm cm}^2$ is the Thomson
scattering cross-section, $n_{e}(z)$ is the mean electron number
density at $z$, and $c$ is the speed of light. The mean electron
number density is given by
\begin{equation}
n_{e}(z)=n_{e}(0)(1+z)^{3} x_{e}(z),
\end{equation}
where $x_{e}$ is the ionization fraction. For a flat $\Lambda$CDM model,
$H(z)=H_0[\Omega_{\Lambda}+\Omega_{\rm m}(1+z)^3]^{1/2}$. 

\begin{inlinefigure}
\resizebox{8cm}{!}{\includegraphics{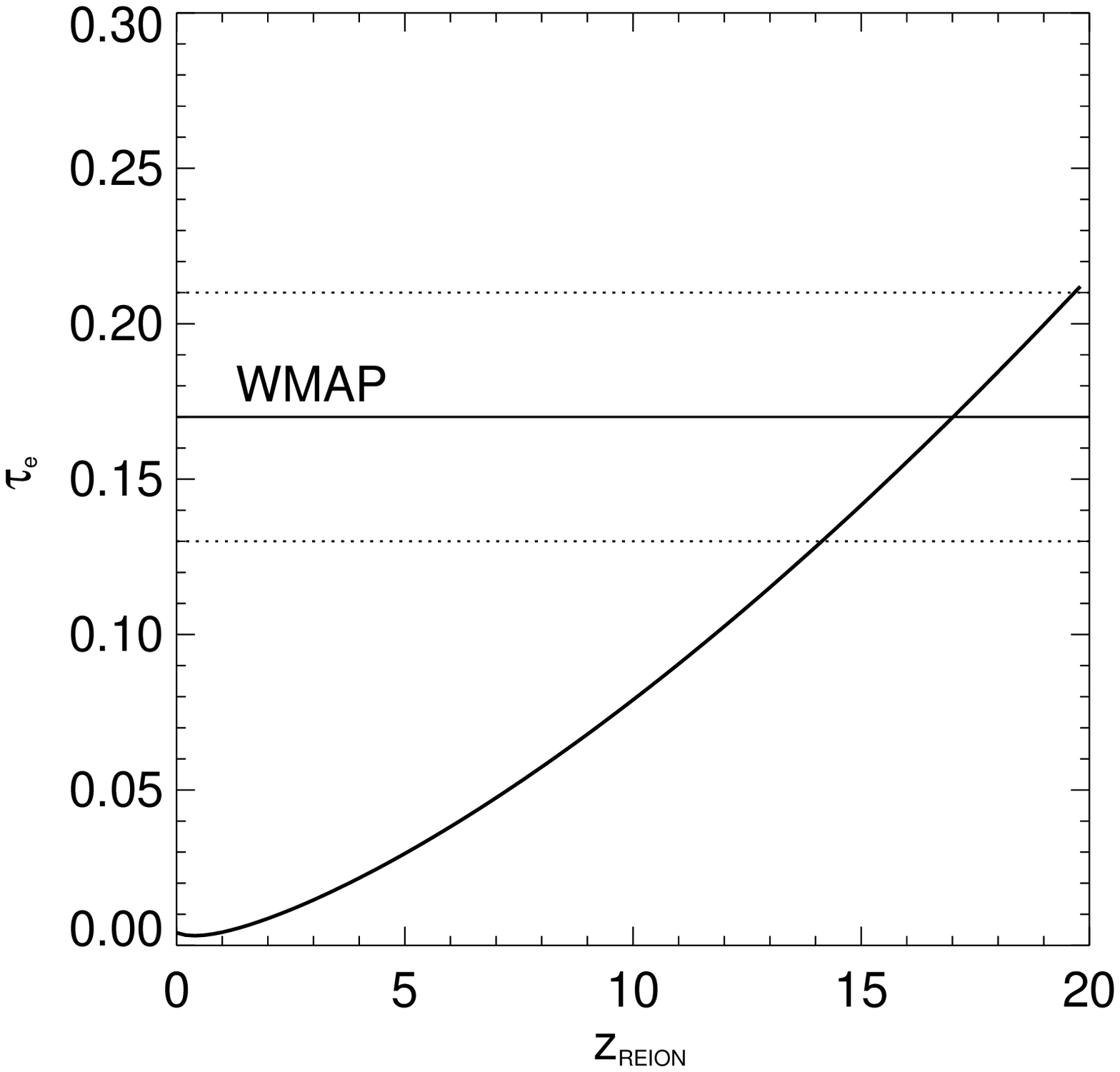}}
\caption{The optical depth to Thomson scattering to $z_{\rm reion}$
when abrupt and complete reionization at $z_{\rm reion}$
is assumed. \label{z_tau}}
\end{inlinefigure}
\\
Putting $x_{e}=1$ for an interval between $z_{1}$ and $z_{2}$, we obtain
the contribution as
\begin{equation}
{\rm d}\tau_{e}|_{z_1\rightarrow z_2}\, =\, {2\over 3}
\frac{\sigma_{\rm T}\,n_{e}(0) c}
{(\Omega_{\rm m} H^{2}_{0})^{1/2}}
\left[
\left ( 1+z \right ) ^3 \, + \, {{\Omega_{\Lambda}}\over{\Omega_{\rm m}}}
\right]^ {1/2} \Bigg | ^{z_2}_{z_1}.
\end{equation}

Consider a simple case where abrupt and complete reionization occurs
at $z$.  Using the WMAP results for the 
cosmological parameters,
$\Omega_{\rm b}h^2$=0.0224 and $\Omega_{\rm m}h^2$=0.135, 
and assuming
a plasma of primordial abundance with, for example,
fully ionized hydrogen and 
singly ionized helium, we obtain the total optical depth to redshift $z$:
\begin{equation}
\int_{0}^{z_{\rm}} {\rm d}\tau_{e} 
\approx 0.0023\left [
\left ( [1+z]^3 \, +\, 2.7\right )^{1/2}-1.93\right ].
\end{equation}

Figure \ref{z_tau} shows the total optical depth against reionization
epoch at $z=z_{\rm reion}$.  The WMAP result is indicated at
$\tau_{e}=0.17$ (solid line) with a 1-$\sigma$ range $0.17\pm0.04$
(dotted line).  The actual reionization history could be more
complicated, but Figure \ref{z_tau} provides the useful insight that a
large Thomson optical depth, say $\tau_{e}>0.13$, can be achieved only
if reionization takes place at an early epoch, $z_{\rm reion}\gtrsim
14$.

Our simulation of early structure formation in the RSI model showed
that there are only 5 gas clouds at $z=15$ in the simulated volume.
As Yoshida et al. (2003b) demonstrate, approximately 100 very massive stars
must be turned on within $\sim$1 recombination time in a (1 Mpc)$^3$
volume to cause complete reionization.  Owing to the strong radiative
feedback from the first stars, the number of stars formed in the
primordial gas clouds is likely to be limited to one (Omukai \& Nishi
1999; Oh et al. 2001).  Hence the number of ``first stars'' in the RSI
model is far too small at $z>15$ to reionize the
Universe.  The epoch of {\it complete} reionization is certainly much
later than $z\sim 15$, implying a small total optical depth $\tau_{e}
\ll 0.13$.

In summary, primordial gas cloud formation which relies on molecular
hydrogen cooling in low-mass, ``mini-halos'' appears to be extremely
inefficient at $z>15$ in the RSI model and thus other sources, such as
early protogalaxies, must contribute to reionization for this model to
be compatible with the WMAP results.  Furthermore, it is important
to establish whether or not ordinary stellar sources in galaxies in
the RSI model can account for reionization early enough to even be
consistent with measurements of the Lyman-$\alpha$ optical depth in
quasars at $z>6$.

\section{Reionization by stars in galaxies}

In the previous section, we have shown that low-mass mini-halos and
primordial gas clouds which can harbor massive, metal free stars (the
``first stars'') do not form in sufficient numbers at high redshift in
the RSI model to account for reionization at $z\simgt 15$.  As we
emphasize in \S 5.2 and Figure 7, reionization must occur early
if the WMAP measurement of the Thomson optical depth is accurate.

The number of larger mass ($\sim 10^7-10^9 M_{\odot}$) halos is also
predicted to be small for the RSI model.  Thus, reionization by stars
in systems that are the likely progenitors of galaxies will be less
efficient than for a $\Lambda$CDM model, for similar assumptions about
the stellar populations.  In this section, we study reionization by
stars in galaxies and proto-galaxies with total masses larger than
$\sim 4\times 10^7 M_{\odot}$, where the gas can cool rapidly by
atomic line transitions.  The nature of the stars forming early in
these environments is uncertain.  It is possible, for example, that
the IMF will be top-heavy, at least initially, perhaps increasing the
production rate of ionizing photons per unit stellar mass compared to
a conventional IMF inferred for stellar populations in the local
Universe.

In what follows, we begin by considering star formation in halos in
the RSI model by assuming that the stellar population is similar to
that locally and is unevolving.  Namely, we adopt a Salpeter (1955)
IMF and normalize the star formation rate (SFR) so that it reproduces
the Kennicutt (1998) law.  For simplicity, we refer to this
description as a ``Population II'' SFR.  In \S 6.4 we examine the
overall effect of relaxing these assumptions.  However, it is already
clear that the RSI model will have difficulty accounting for the WMAP
estimate of the Thomson for any reasonable assumptions about star
formation simply because structure forms too late in this scenario to
produce a large $\tau_e$, as suggested by Figure 7.

To study how reionization proceeds in the RSI model, we run
multi-source radiative transfer simulations using the technique of
Sokasian et al. (2001, 2003a,b).  
In order to locate radiation sources and to compute the IGM density
field, 
we utilize existing outputs of a
large volume $\Lambda$CDM simulation from Springel \& Hernquist (2003b),
normalizing the star formation rate expected for the RSI model using
the analytic description of Hernquist \& Springel (2003), which can be
applied to any cosmology.

\subsection{The star-formation rate in the RSI model}

First, we compute the ``generic'' star-formation rate for the RSI
model using the analytic model of Hernquist \& Springel (2003).  Their
approach describes the formation of dark matter halos using an analytic
halo mass function (Sheth \& Tormen 1999; 2002), and employs
simplified prescriptions for the cooling of gas and star-formation
which are commonly used in semi-analytic models of galaxy formation
(Kauffmann et al. 1999; Springel et al. 2001; Yoshida et al. 2002).
This analytic formalism has been tested in detail by comparing its
predictions to those from

\begin{inlinefigure}
\resizebox{8cm}{!}{\includegraphics{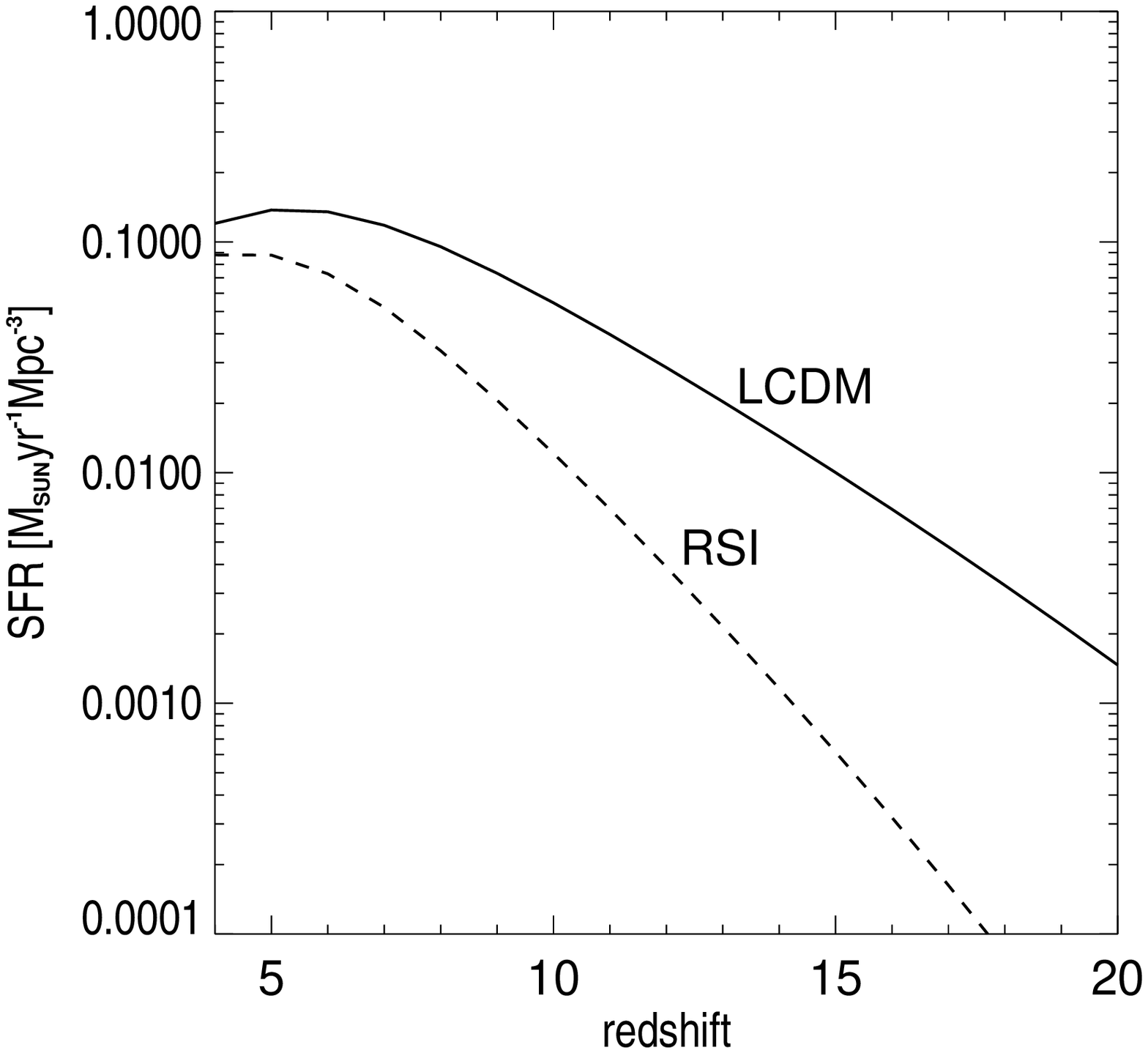}}
\caption{The star-formation rate density in galaxies 
(``Population II'') as a function of redshift
for the $\Lambda$CDM (solid line) and RSI (dashed line) models.  
\label{sfr}}
\end{inlinefigure}
\\
cosmological simulations that include
hydrodynamics and star formation (Springel \& Hernquist 2003b).  A
thorough description of the model is given by Hernquist \& Springel
(2003) and references therein.

In particular, Hernquist \& Springel (2003) show that under rather
general circumstances the star-formation rate 
density is well-approximated by the simple form
\begin{equation}
\dot{\rho}_{*}(z)=\dot{\rho}_{*}(0)\frac{\chi^{p}}{1+\alpha
(\chi-1)^{q} \exp(\beta \chi^{r})},
\label{hs_sfr}
\end{equation}
where
\begin{equation}
\chi(z) \equiv \left ( {{H(z)}\over{H_0}} \right ) ^{2/3} \,\,.
\end{equation}
In equation (\ref{hs_sfr}), the factors $\beta$ and $r$ are fixed by
the underlying cosmological power spectrum, the choices for
$p$ and $q$ reflect the interplay between radiative
cooling in halos and the expansion of the Universe, and the
normalizing constants $\dot{\rho}_{*}(0)$ and $\alpha$ depend
on the rate at which cold, dense gas is converted into stars.

The form of equation (\ref{hs_sfr}) is motivated by two competing
effects which regulate star formation.  At sufficiently high
redshifts, collapsed gas is dense and cooling times are
short so that star formation is only limited by the gravitational
growth of massive halos.  At low redshifts, the cooling time in
halos increases owing to the decline in the mean density of the
Universe by cosmic expansion.  In this regime, the supply of
star-forming gas is throttled by the expansion of the Universe.

Based on the multiphase description of star-forming gas described 
by Springel \& Hernquist (2003a), Hernquist \& Springel
(2003) show that the appropriate values for the parameters in
equation (\ref{hs_sfr}) for the $\Lambda$CDM model employed here are
$p=2$, $q=3$, $r=7/4$, $\alpha=0.012$, $\beta=0.041$, and
$\dot{\rho}_{*}(0)=0.013 M_\odot$/yr/Mpc$^3$. 

For the RSI cosmology, we use the RSI linear power spectrum as input 
and follow the procedure described in Hernquist \& Springel (2003).
We find that the resulting SFR density is again well-fitted by
equation (\ref{hs_sfr}) for the choices 
$p=1.5$, $q=3.05$, $r=1.85$, $\alpha=0.01$, $\beta=0.057$, and
$\dot{\rho}_{*}(0)=0.0176 M_\odot$/yr/Mpc$^3$. 

In Figure \ref{sfr}, we compare the SFR density estimated in this
manner for the two cosmologies.  The RSI model predicts a
significantly smaller SFR than the $\Lambda$CDM model, in qualitative
agreement with the semi-analytic estimates of Somerville et
al. (2003).  We use the forms for the
SFR density in our ray-tracing calculatiosn 
only within the restricted redshift range $5<z<20$ to compute
radiation source luminosities.

\subsection{Source luminosities}

The second step is to prescribe the luminosity of the sources
(``galaxies'') located in the simulations.  Specifically, we use the
outputs from Run Q5 of Springel \& Hernquist (2003b), which simulates
a cube of 14.3 Mpc on a side.  We first calibrate the SFRs of
individual sources identified in the simulation, and then convert
these into luminosities.  As in Sokasian et al. (2003a), we use the
derived SFR (equation \ref{hs_sfr}) as a reference to match the total
SFR of the sources.  We carry out the calibration procedure after
selecting sources in the simulation outputs in the following two
ways.

\subsubsection{Dim source model}

In this approach, we identify all the halos found in each output as
``galaxies'' and assign luminosities to them based on the actual
star-formation rates which are computed in a dynamically consistent
way in the cosmological simulation.  The minimum mass of the sources
is set by the resolution of the simulation, which is $M_{\rm min}=4.3
\times 10^7 h^{-1}M_{\odot}$ in our case.  The SFRs of the sources
with masses close to the resolution limit are corrected in the same
manner as in Sokasian et al. (2003a, \S 4.2.1).  We then scale the
SFRs of all the galaxies by multiplying by a constant factor $C_{\rm
L} (z)$ such that the total SFR within the simulated volume matches
that for the RSI model obtained from equation (\ref{hs_sfr}).  Note
that the scaling factor $C_{\rm L} (z)$ is kept constant for all the
galaxies within one simulation output, but it is time-dependent.

\subsubsection{Massive halo model}

Since the halo abundance itself is lower in the RSI model than in
the $\Lambda$CDM cosmology, it may be more appropriate to populate a
smaller number of objects as radiation sources in the RSI model.
Motivated by this, we first compute the number density of halos with
virial temperature greater than $10^4$ K in which the gas can cool by
atomic line transitions.  The threshold mass is given by
\begin{equation}
M_{\rm min} = 10^8 \left(\frac{1+z}{11}\right)^{-3/2}\;\;M_{\odot}.
\end{equation}
Then we use the halo mass function to compute the expected abundance
of halos with mass greater than $M_{\rm min}$.  Let us denote this
number by $N_{\rm source} (z)$.  In the simulation output produced at
$z$, we select $N_{\rm source} (z)$ halos in order of their
mass.  Luminosities are assigned to the selected sources in proportion
to their mass, $L \propto M_{\rm halo}$ such that the total SFR summed
over the selected sources matches the global SFR given by equation
(\ref{hs_sfr}), with the parameter values appropriate for the 
RSI model.

\subsubsection{Escape fraction}

Finally, the ionizing flux from the sources is modulated by the
``escape fraction,'' $f_{\rm esc}$, which describes the probability
that individual photons can escape from the galaxy in which they were
produced into the surrounding intergalactic medium (IGM).
Sokasian et al. (2003a) studied reionization by stellar sources
in the $\Lambda$CDM universe by adopting several values for 
$f_{\rm esc}$.  Based on a detailed
comparison with the observational results of the $z=6.28$ quasar (SDSS
1030.10+0524) in Becker et al. (2001), 
they conclude that $f_{\rm esc}=0.2$ is most
consistent with the observed Lyman-$\alpha$ flux transmittance.  We
take this $f_{\rm esc}=0.2$ model in Sokasian et al. (2003a) as our
fiducial case for the $\Lambda$CDM model.  For the RSI model, our
zeroth order requirement is that reionization should be completed by
$z=7$. After several experiments, we found that setting $f_{\rm
esc}>0.6$ is necessary for the RSI model to satisfy this condition.
We thus set $f_{\rm esc}=0.6$ as our basic choice for the RSI model and
compare the results with the fiducial $\Lambda$CDM model.

\subsection{Numerical results for ``Population II'' star formation}

We compute the total volume filling factor of the ionized medium
$Q_{\rm HII}$ from the results of our ray-tracing calculations.
Figure \ref{QHII} shows the resulting ionized volume fraction, and
Figure \ref{tot_tau} shows the Thomson optical depth $\tau_{e}$ as a
function of redshift.  In the figures, we show only the results for
our ``massive-halo model'' of source selection for the RSI model.
Comparing the results for the dim-source model and the massive-halo
model, we found that there is no substantial difference between the
two in either $\tau_{e} (z)$ or $Q_{\rm HII} (z)$.  We note, however,
that the overall morphology of the ionized regions looks somewhat
different between the two cases, because there are more sources in the
dim-source model that reside in less dense regions.  Nevertheless, for
global quantities such as the total optical depth, the two models are
found to be quite similar.

On the other hand, the results for the $\Lambda$CDM and the RSI models
differ substantially. As Figure \ref{QHII} shows, the ionized volume
fraction in the RSI model remains
small until $z\sim 9$, and then rapidly increases close to unity at
$z\sim 7$.  At later epochs, $z\lesssim 9$, the abundance of high
mass ($\sim 10^9 M_{\odot}$) halos rapidly increases, and the total
star-formation (and hence the photon production) is dominated by such
large halos.  This can also be understood 
in terms of the halo mass function at $z=7$
shown in Figure \ref{mfz7}.  By this time, the mass functions for the
two models are already quite similar, with the difference being only a
constant factor of roughly two (compare with the much larger difference at
$z=17$ at smaller mass scales shown in Figure \ref{mfz17}).  Namely,
the RSI model quickly catches up with the $\Lambda$CDM model in the
abundance of halos around $z\sim 7-11$ (see also the increasing SFR in
Figure \ref{sfr}), which causes the rapid increase in the ionized
volume fraction as seen in Figure \ref{QHII}.  Reionization is nearly
completed ($Q_{\rm HII}>0.95$) at $z = 8$ in the $\Lambda$CDM model,
whereas it occurs at $z\approx 7.3$ in the RSI model.  However, as shown in
Figure \ref{tot_tau}, the resulting total optical depth
$\tau_{e}=0.055$ for the RSI model is still small, owing to the small
$Q_{\rm HII}$ at earlier epochs.  We emphasize that a large Thomson
optical depth can be achieved only if reionization occurs early
(\S 6).  In other words, after reionization is completed, the
additional

\begin{inlinefigure}
\resizebox{8cm}{!}{\includegraphics{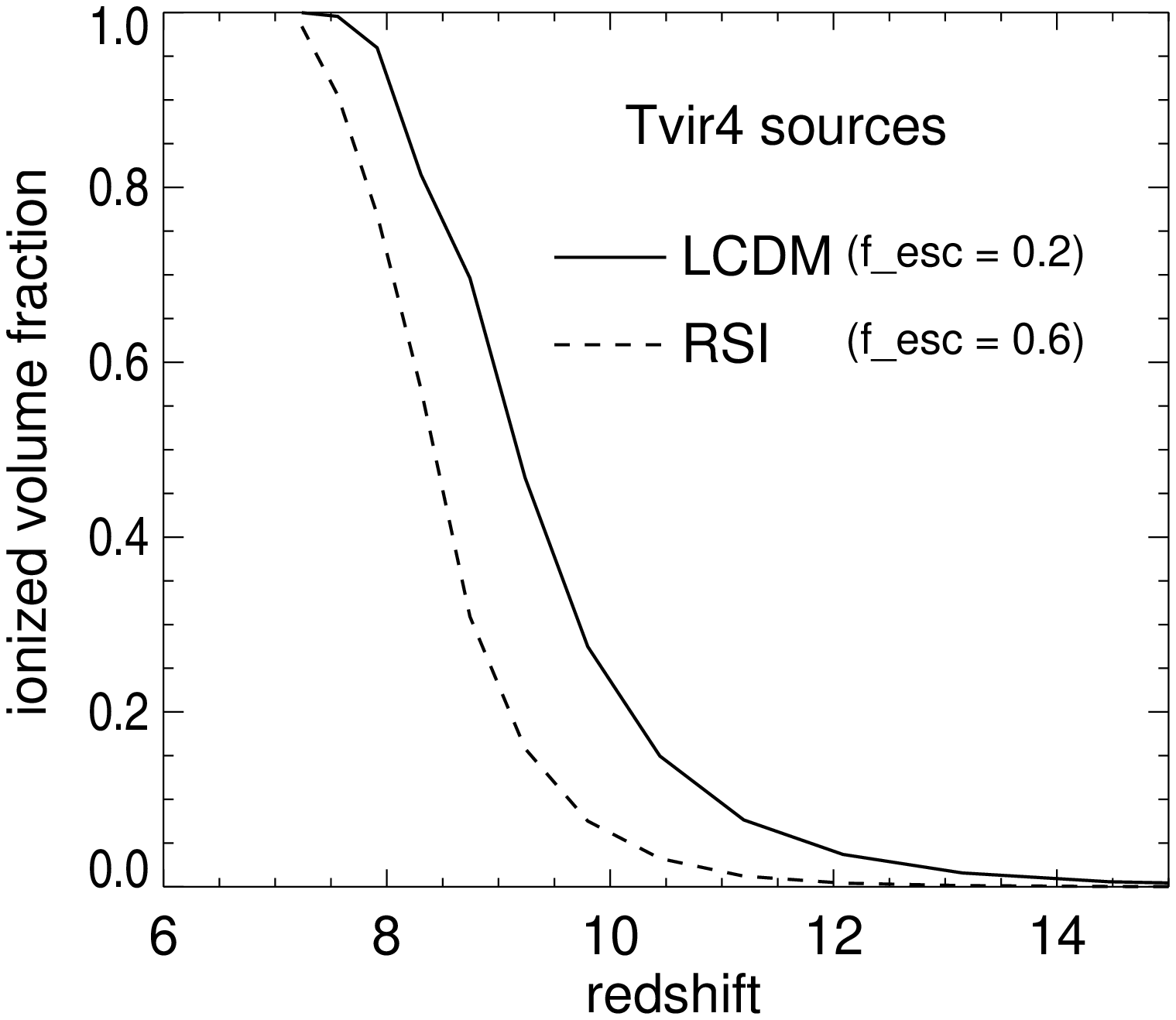}}
\caption{The ionized volume fraction for the $\Lambda$CDM model 
with $f_{\rm esc}=0.2$ (solid line) and for
the RSI model with $f_{\rm esc}=0.6$ (dashed line),
for ``Population II'' stars. \label{QHII}}
\end{inlinefigure}
\begin{inlinefigure}
\resizebox{8cm}{!}{\includegraphics{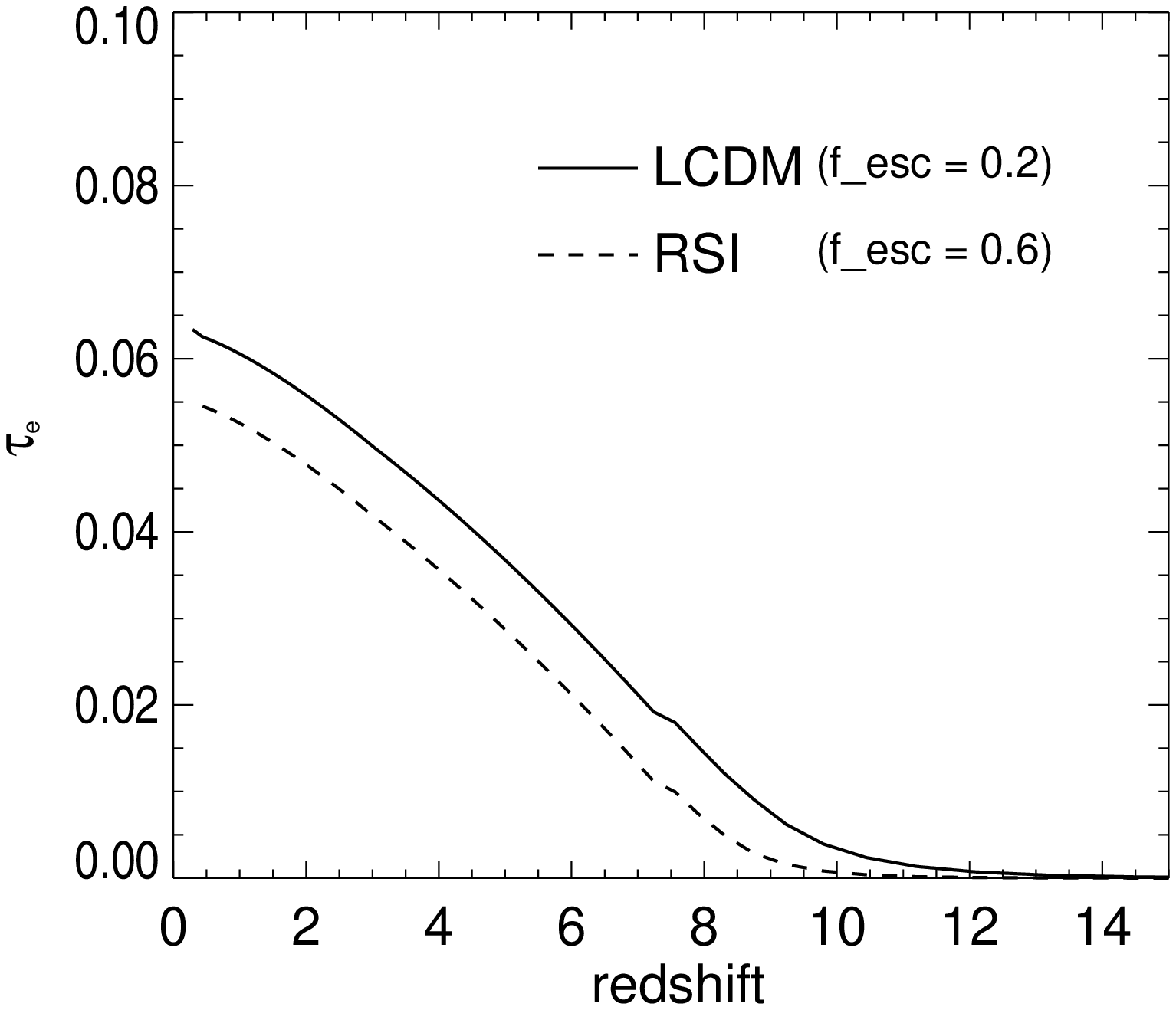}}
\caption{The Thomson optical depth computed from the outputs 
of our simulations for the $\Lambda$CDM model 
with $f_{\rm esc}=0.2$ (solid line) and for
the RSI model with $f_{\rm esc}=0.6$ (dashed line) for ``Population
II'' stars. \label{tot_tau}}
\end{inlinefigure}
\\
photon supply does not matter in the
total optical depth as long as the IGM is kept almost fully ionized
thereafter, as implied by Figure 7.
(Note that here we discuss only hydrogen reionization.)

Our fiducial $\Lambda$CDM case with $f_{\rm esc}=0.2$ predicts a
larger total optical depth than the RSI model with $f_{\rm esc}=0.6$
by $\Delta \tau_{e}\sim 0.01$.  The difference mostly comes from the
fact that $Q_{\rm HII}$ is still quite small at $z>9$ in the RSI
model.  Interestingly, none of the ``conventional'' scenarios we have
considered for ``Population II'' star formation reproduces the
claimed high optical depth from WMAP of $0.13<\tau_{e}<0.21$
(1-$\sigma$; Kogut et al. 2003).
It is clear that these models are inconsistent with the WMAP measurement
and thus may need, for example, additional radiation sources to ionize 
the IGM early on.

In related work, we have shown that an early generation of massive,
metal-free stars in mini-halos supplemented by ``ordinary'' star
formation at lower redshifts to account for the spectra of high
redshift quasars can yield a Thomson optical depth approaching that of
the WMAP measurement in a $\Lambda$CDM cosmology (Yoshida et
al. 2003b, 

\begin{inlinefigure}
\resizebox{8cm}{!}{\includegraphics{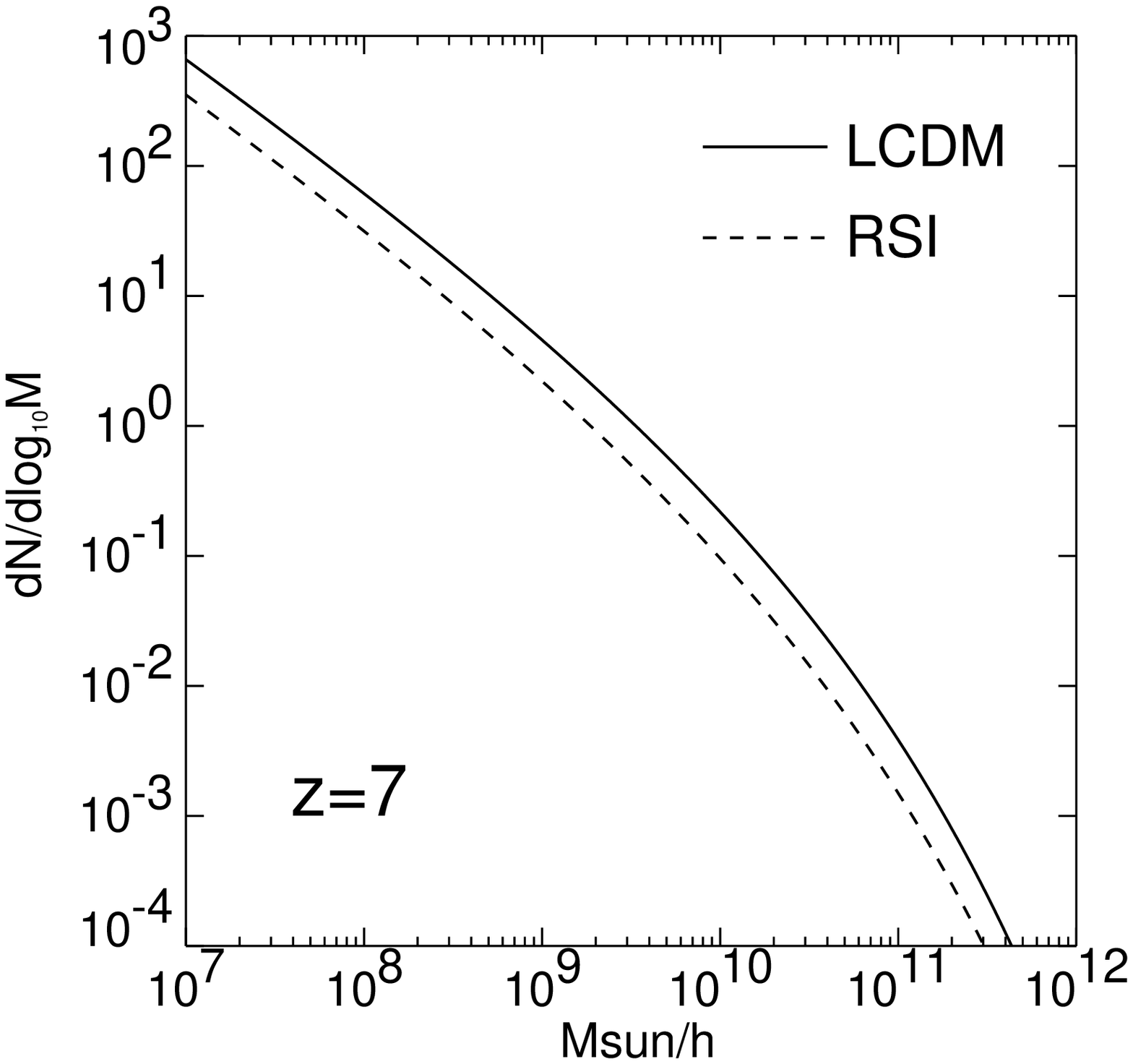}}
\caption{The Press-Schechter mass function at $z=7$.\label{mfz7}}
\end{inlinefigure}
\\
Sokasian et al. 2003b), along the lines suggested by Haiman
\& Holder (2003) and Cen (2003).  However, it does not appear that a
similar conclusion will obtain for the RSI model because of the low
abundance of mini-halos at high redshifts, as we found in the case of
a warm dark matter model with particle mass 10 keV (Yoshida et
al. 2003b).

\subsection{Star formation in galaxies with boosted photon emission}

The results described above indicate that the RSI model is
incompatible with a high Thomson optical depth like that suggested by
the WMAP measurements.  A remaining loophole in this
argument is that the stellar
population in galaxies at early times may have differed from that in our
underlying model so that a larger number of ionizing photons were
produced from galaxies at $z>10$ than we assumed
above (e.g. Ciardi, Ferrara \& White 2003;
Wyithe \& Loeb 2003; Sokasian et al. 2003a).
We first discuss such possibilities quantitatively, and then
show numerical results for a model that enhances the photon 
production rate in galaxies. 
 
For a Salpeter-type power-law IMF, the
hydrogen ionizing photon emission rate of zero-metallicity stars is
larger than that of a Population II (i.e. having metals) counterpart
only by about 50\%
(Thumlinson \& Shull 2000; see Leitherer et al. 1999 for
low-metallicity cases.). The stellar IMF in the first galaxies 
may have had even an extreme top-heavy shape, with massive stars exclusively 
forming out of a primeval gas
(Schwarzschild \& Spitzer 1953; Matsuda, Sato \& Takeda 1969; 
Yoshii \& Saio 1986; Larson 1998; 
Abel, Bryan \& Norman 2002; Bromm, Coppi \& Larson 2002).
Very massive Population III stars
are efficient emitters of UV photons, with the hydrogen ionizing photon
emission rate per stellar mass being up to 10-20 times higher than for
ordinary stellar populations (Bromm, Kudritzki \& Loeb 2001; Schaerer 2002).
It is thus conceivable that, in early galaxies, the photon emission
rate per 
stellar mass was $\sim 10$
times higher than that of the ``Population II'' case
we considered above. 

More specifically, the predicted 
star-formation rate for the RSI model shown in Figure \ref{sfr} is 

\begin{inlinefigure}
\resizebox{8cm}{!}{\includegraphics{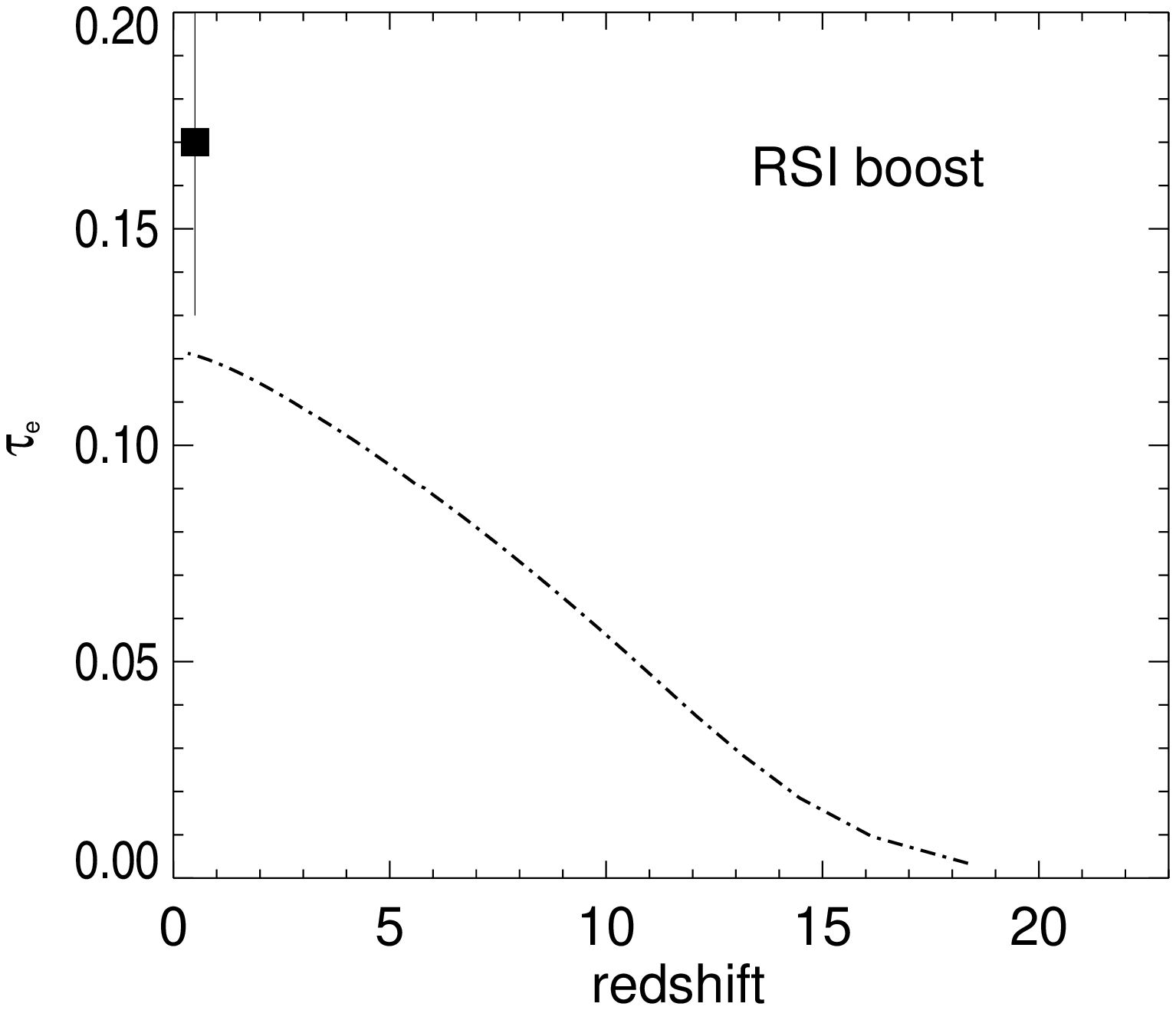}}
\caption{The Thomson optical depth for the RSI model in which 
the photon emission rate is ``boosted'' according to equation (\ref{eq_boost}). 
The filled square is 
the WMAP result with its 1-$\sigma$ error bar.\label{tau_boost}}
\end{inlinefigure}
\\
$\sim 5\times 10^{-4} M_{\odot} {\rm yr}^{-1} {\rm Mpc}^{-3}$ at
$z=15$. Assuming a constant conversion factor of $10^{53} {\rm
s}^{-1}$ ionizing photons per star-formation rate of 1 $M_{\odot}{\rm
yr}^{-1}$ for an ordinary stellar population with solar metallicity
(Madau, Haardt \& Rees 1999), we obtain a total rate of
$\sim 5\times 10^{49}$ ionizing photons per second per cubic comoving 
1 Mpc volume at $z=15$ for our fiducial RSI model. 
If all the formed stars have a mass of $300M_{\odot}$,
the star-formation rate of $\sim 5\times 10^{-4} M_{\odot} {\rm yr}^{-1} {\rm Mpc}^{-3}$
corresponds to the formation of 5 such stars every three million years
within the volume. The calculation of Bromm et al. (2001)
suggests that a total emission rate of $2.5\times 10^{51}$
ionizing photons per second, a 50 times higher photon emission rate, 
can be maintained in this case, {\it assuming} such very massive
stars are continuously formed.
(Note that these massive stars are typically short-lived,
with a main-sequence lifetime of about two million years.) 
We thus conclude that a 100 times higher photon production rate 
than assumed in our original RSI model serves as a 
conservative upper limit.

We examine the overall effect of such an additional photon supply
by ``boosting'' the photon emission rate of the sources in the RSI massive-halo model.
Sokasian et al. (2003a) found that boosting the photon emission rate 
at high redshifts progressively shifts the reionization epoch (see their Figure 13).
Motivated by their result, we employ a simple prescription which increases the emission
rate of individual sources according to
\begin{equation}
\dot{N}_{\rm ph, eff.}=\exp(0.53\;(z-6)) \times \dot{N}_{\rm ph},
\label{eq_boost}
\end{equation} 
where $\dot{N}_{\rm ph}$ and $\dot{N}_{\rm ph, eff.}$ are the
original and the boosted photon emission rates, respectively.
With this parametrization, the ionizing photon emission rate
of the galaxies steeply increases toward higher redshift, modeling
either a transition of the stellar IMF or a 
contribution from massive Population III stars or both.
This rather extreme model requires effectively 
more than a $10$ times higher photon emission rate at $z>10$,
and more than $100$ times higher at $z>15$.
We carry out radiative transfer simulations for the RSI model with
this luminosity ``boost''. 
Figure \ref{tau_boost} shows the resulting $\tau_{e} (z)$.
The reionization epoch is found to be shifted to earlier times, $z\sim 10$, 
than in the case considered in \S 6.3,
and consequently $\tau_{e}$ is greatly increased compared with the 
fiducial RSI model. However, the total optical depth for this boost model
reaches $\tau_{e}\sim 0.12$ at $z=0$, {\it still lower} 
than the 1-$\sigma$ range of the WMAP result.  

\section{Summary}

Our numerical simulations of early structure formation show that
primordial gas cloud formation in low-mass ``mini-halos'' is very
inefficient at $z>15$ in the RSI model, making it unlikely that the
``first stars'' contribute significantly to reionization in this
scenario.  Using radiative transfer calculations, we also show that
reionization by ordinary (``Population II'') stellar sources in
galaxies is completed late in the RSI model if we employ conventional
models of star-formation.  In order to be compatible with the epoch of
the end stage of reionization inferred from high redshift quasar
observations, the photon escape fraction from galaxies must be large,
$f_{\rm esc}\simgt 0.6$.  We found that the resulting total Thomson
optical depth in this case is $\tau_{e}\approx 0.055$, in apparent
conflict with the recent WMAP measurements of CMB polarization.  

In order for the total optical depth to be as large as $\tau_{e}>0.1$,
reionization must be completed before $z \sim 10$ (Figure
\ref{z_tau}).  Ciardi et al. (2003) and Wyithe \& Loeb (2003) suggest
that employing an effectively top-heavy IMF may be necessary to cause
early reionization.  Cen (2003) argues that a large Thomson optical
depth could be obtained if the ``Population III era'' is prolonged so
that massive stars continuously form until quite low redshift.  If a
large volume of the IGM could be kept chemically pristine until low
redshift ($z\sim 10$) due to the very low global star-formation rate
in the RSI model, the stellar IMF in galaxies may be top-heavy
(Schneider et al. 2002; Mackey, Bromm \& Hernquist 2003).
Unfortunately, little is 
presently known about star-formation in a primordial
gas in high mass halos --the ``first galaxies''-- and clearly needs
to be addressed using numerical simulations as well as theoretical
modeling (Larson 1998; Omukai 2001; Oh \& Haiman 2002).  We have
explored the possibility of enhancing the photon emission rate by
employing a simple prescription.  The model parameters were chosen
such that the photon emission rate is close to a 
plausible maximum bound.
Our analysis has shown that, whereas the extreme boost model does predict an
earlier reionization epoch $z_{\rm reion} \sim 10$ and a larger total
optical depth $\tau_{e} \sim 0.12$, the result is still marginally
inconsistent with the WMAP data.

Recent numerical simulations by Bromm, Yoshida \& Hernquist (2003)
show that early Population III supernovae quickly pollute the
surrounding IGM with metals.  If many massive population III stars are
formed in ``mini-halos'' and explode as hyper-energetic supernovae,
the global metal-enrichment of the IGM is quickly achieved to such a
degree that a transition of the stellar IMF is caused at a very early epoch
(Yoshida, Bromm \& Hernquist 2003).  Then, the large photon production
rate at $z>6$ we hypothesized as coming from metal-free stars cannot
be maintained for a sufficiently long time, and the reionization
history in the RSI model will be just as shown in Figure \ref{QHII},
with a small total optical depth ($\tau_{e} <0.06$) as shown in Figure
\ref{tot_tau}.
  
In the end, we are left with a contradiction: the running of the
primordial power spectrum favored by the WMAPext + 2dF +
Lyman-$\alpha$ analysis and the high Thomson optical depth measured by
the WMAP satellite appear to be inconsistent with one other.  We have
discussed various possibilities to resolve the conflict, and conclude
that a rather radical solution or a fine-tuned combination, if any is
possible, of the proposed resolutions are necessary.  Models with a
weaker running of the primordial power spectrum will alleviate the
discrepancy slightly, but the overall results would not be changed.
If the high Thomson optical depth is confirmed and the running 
of the primordial power spectrum turns out to be real, 
it may be necessary to invoke exotic radiation sources other than 
stellar populations or some ionization mechanisms.
Such scenarios include decaying particles (Sciama 1982;
Dodelson \& Jubas 1992; Hansen \& Haiman 2003)
or formation of early mini-quasars (Eisenstein \& Loeb 1995;
Sasaki \& Umemura 1996).

The prospects for observationally resolving this discrepancy in the near
future appear bright.
Data from planned CMB polarization experiments, by the continued 
operation of WMAP, and post-WMAP observatories such as Planck will pin 
down a precise value for $\tau_{e}$.  Lyman-$\alpha$ forest
observations exploiting a large sample of SDSS quasars will place a
tighter constraint on the matter power spectrum on large scales,
as will the SDSS galaxy redshift survey.
In the longer term, it may be possible to map out the {\it evolution}
of reionization from redshifted 21 cm emission using instruments
such as the Square Kilometer Array\footnote{http://www.usska.org}
or the Low Frequency Array\footnote{http://www.lofar.org}
(e.g. Madau et al. 1997; Tozzi et al. 2000; Iliev et al. 2002;
Ciardi \& Madau 2003).  For example, Furlanetto et al. (2003)
show that frequency fluctuations can be used to distinguish between
models with a different number of reionization epochs.
Analyses
of these various high-precision data promise to provide a more complete 
picture of
the matter density distribution in the early Universe over a wide
range of scales and its relationship to the formation of stars and
galaxies.
\\

We thank James Bullock, Volker Bromm, and Ravi Sheth for helpful
comments on the earlier draft.  NY acknowledges support from 
the Japan
Society of Promotion of Science Special Research Fellowship.  This
work was supported in part by NSF grants ACI 96-19019, AST 98-02568,
AST 99-00877, and AST 00-71019.  The simulations were performed at the
Center for Parallel Astrophysical Computing at the Harvard-Smithsonian
Center for Astrophysics.

\end{document}